\documentclass[11pt]{article}

\usepackage[final]{acl}

\usepackage{times}
\usepackage{latexsym}

\usepackage[T1]{fontenc}
\usepackage[utf8]{inputenc}

\usepackage{microtype}

\usepackage{inconsolata}

\usepackage{graphicx}

\usepackage{booktabs}
\usepackage{array}
\usepackage{amsmath}
\usepackage{longtable}
\usepackage{enumitem}
\usepackage{multirow}
\usepackage{makecell}
\usepackage{subcaption}
\usepackage{epstopdf}
\usepackage{xcolor}
\usepackage{tcolorbox}
\usepackage{mdframed}
\usepackage{tabularx}
\usepackage{longtable}
\usepackage{tabularray}
\usepackage{stfloats}
\usepackage{soul}
\usepackage[table]{xcolor}
\tcbuselibrary{listings, breakable, skins}
\usepackage{adjustbox}

\usepackage[vlined, linesnumbered, inoutnumbered, ruled]{algorithm2e}
\usepackage[misc]{ifsym}

\newlist{userlist}{itemize}{2}
\setlist[userlist]{label=-, leftmargin=2em}
\setlist[userlist,1]{label=-} % 第一级列表样式
\setlist[userlist,2]{label=-, leftmargin=4em} % 第二级列表缩进

\title{TagRAG: Tag-guided Hierarchical Knowledge Graph Retrieval-Augmented Generation}

\author{
 \textbf{Wenbiao Tao},
 \textbf{Xinyuan Li},
 \textbf{Yunshi Lan\thanks{Corresponding author}},
 \textbf{Weining Qian}
\\
 East China Normal University
\\
 \{wbtao, xyli\}@stu.ecnu.edu.cn, \{yslan, wnqian\}@dase.ecnu.edu.cn
}

\begin{document}
\maketitle

\begin{abstract}
Retrieval-Augmented Generation enhances language models by retrieving external knowledge to support informed and grounded responses. However, traditional RAG methods rely on fragment-level retrieval, limiting their ability to address query-focused summarization queries. GraphRAG introduces a graph-based paradigm for global knowledge reasoning, yet suffers from inefficiencies in information extraction, costly resource consumption, and poor adaptability to incremental updates. To overcome these limitations, we propose TagRAG, a tag-guided hierarchical knowledge graph RAG framework designed for efficient global reasoning and scalable graph maintenance. TagRAG introduces two key components: (1) Tag Knowledge Graph Construction, which extracts object tags and their relationships from documents and organizes them into hierarchical domain tag chains for structured knowledge representation, and (2) Tag-Guided Retrieval-Augmented Generation, which retrieves domain-centric tag chains to localize and synthesize relevant knowledge during inference. This design significantly adapts to smaller language models, improves retrieval granularity, and supports efficient knowledge increment. Extensive experiments on UltraDomain datasets spanning Agriculture, Computer Science, Law, and cross-domain settings demonstrate that TagRAG achieves an average winning rate of 78.36\% against baselines while maintaining about 14.6x construction and 1.9x retrieval efficiency compared with GraphRAG.
\end{abstract}

\section{Introduction}
Retrieval-Augmented Generation (RAG)~\cite{lewis2020retrieval} is a framework that enhances the output of language models by retrieving relevant documents from an external knowledge source and conditioning the generation process on both the input query and the retrieved content, enabling more informed and factually grounded responses~\cite{fan2024survey, chen2025comrag}. For Large Language Models (LLMs), RAG has become the most important technology to help them land effectively in the fields of medicine~\cite{zhao2025medrag}, law~\cite{wiratunga2024cbr}, finance~\cite{barry2025graphrag}, education~\cite{lan2025survey}, etc.
% However, constrained by its fragment-oriented retrieval, the RAG framework focuses on matching information within a limited span, leading to its lack of a global view to cope with highly abstract questions~\cite{peng2024graph}.
However, conventional RAG approaches typically rely on unstructured text retrieval, which often fails to capture the intricate semantic relationships required for complex reasoning~\cite{liang2024aligning}, thereby motivating the integration of graph-based question answering~\cite{tao2024finqa} to better leverage structured knowledge~\cite{zhu2025graph}.

\begin{figure}[t]
 \centering
 \includegraphics[width=0.98\linewidth]{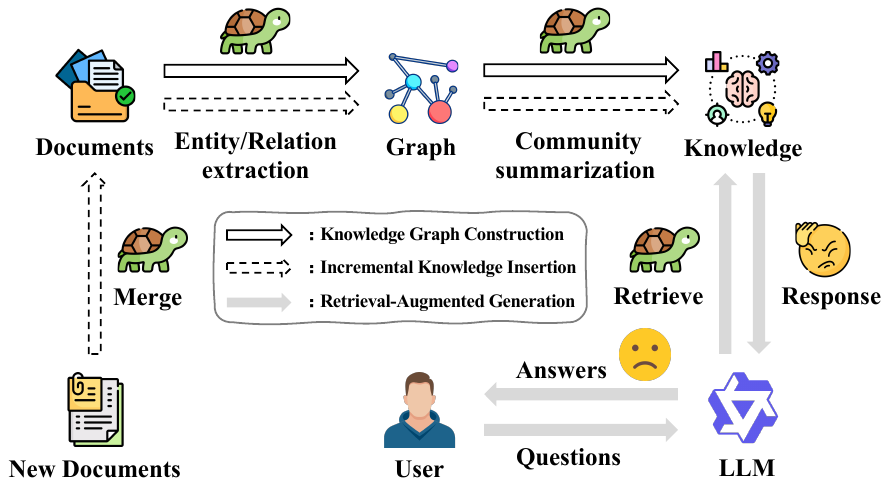}
 \centering
 \caption{Inefficient graph construction and reasoning.}
 \label{fig:motivation}
\end{figure}

The emergence of GraphRAG~\cite{edge2024local} solves this problem by extracting entities from documents, dividing knowledge communities, and generating community summaries, thereby refining global information. The new paradigm of introducing knowledge summarization into RAG at the graph level makes it realistic for LLMs to generate responses from a global view.
However, GraphRAG still suffers from the drawbacks of inefficient information extraction and expensive resource calls. 
To solve these problems, some methods, such as LightRAG~\cite{guo2024lightrag} and KET-RAG~\cite{huang2025ket}, simplify the knowledge graph structure, avoiding complicated community divisions and reducing construction costs significantly.
Besides, for improving the efficiency of inference, methods like MiniRAG~\cite{fan2025minirag}, FG-RAG~\cite{hong2025fg} and LeanRAG~\cite{zhang2026leanrag} retrieve relevant subgraphs to generate query-aware fine-grained answers.
However, these approaches are divorced from a global perspective. Although it is possible to inject the intrinsic knowledge of LLMs into the entity in the process of building the graph, it is difficult to compensate for the comprehensive understanding of the complete resource library. 

Based on the above issues, we revisit GraphRAG and present the following two research questions.

% For enriching retrieved information, the hypergraph~\cite{luo2025hypergraphrag}, the hierarchical architecture~\cite{hong2025fg} and the query-level fine-grained summarization~\cite{huang2025retrieval} are utilized to enhance knowledge representation.
% Relying on the inference capabilities of powerful LLMs such as GPT-4o, these frameworks invoke expensive resources on a large scale, making it difficult to deploy to low-resource scenarios such as edge computing.
% On the other hand, for improving the efficiency of graph construction, LightRAG~\cite{guo2024lightrag}, MiniRAG~\cite{fan2025minirag} and KET-RAG~\cite{huang2025ket} replace community-level summaries with entity-level abstract descriptions, thus reducing indexing costs.
% However, these approaches are divorced from a global perspective. Although it is possible to inject the intrinsic knowledge of LLMs into the entity in the process of building the graph, it is difficult to compensate for the comprehensive understanding of the complete resource library.

% These GraphRAG-based methods provide a retrieval environment for RAG by constructing knowledge graphs and indexing vector libraries.
% However, by extracting entities and relations from documents, all the knowledge they extract is embedded in flat-level units, which is obviously unfavorable for managing such a large knowledge structure.
% Therefore, we revisit Graph RAG and consider how far it is from the actual landing.
% In light of the shortcomings of the existing methods mentioned above, we summarize the following two research questions.

\begin{itemize}
    \item \textbf{(RQ1) How to enhance the global knowledge graph representation and reasoning with low resource consumption?} Existing Graph RAG methods heavily rely on large-scale LLMs (e.g., GPT-4o~\cite{hurst2024gpt}) to handle construction, retrieval, and generation, hindering migration to smaller models due to limited capacity. These high-intensity model calls also incur substantial resource costs and constrain deployment flexibility. Therefore, decoupling LLMs' end-to-end dominance becomes essential for practical and scalable applications.
    \item \textbf{(RQ2) How to achieve robust and efficient knowledge increment for large-scale Graph RAG?} GraphRAG employs a bottom-up construction paradigm that involves entity/relationship extraction followed by community summary generation. However, this architecture struggles with incremental updates, requiring costly full-graph reconstruction. Consequently, it is crucial to develop an efficient incremental construction mechanism that minimizes overhead and integrates seamlessly with existing knowledge structures.
\end{itemize}

To tackle these challenges, we propose TagRAG, a tag-guided hierarchical knowledge graph retrieval-augmented generation framework, comprising Tag Knowledge Graph Construction and Tag-guided Retrieval-Augmented Generation. 
In the construction stage, TagRAG constructs a Tag Knowledge Graph by extracting object tag keywords and their relationships from documents. These object tags are linked to predefined root domain tags, forming hierarchical domain tag chains. Knowledge from object tags connected to domain tags is summarized and fused for information integration.
During inference, domain-centric tags are retrieved to localize knowledge. The hierarchical tag chains are collected to integrate synthesized information, which is then fused into global answers via tag-guided retrieval-augmented generation.
In summary, the contributions are as follows:

\begin{itemize}
    \item We propose TagRAG, a powerful and efficient graph-based RAG framework that enhances global knowledge representation and reasoning in low-resource scenarios.
    \item We introduce the architecture of domain tag chains, which improves the hierarchy of knowledge graphs and optimizes the adaptability of incremental insertion.
    \item We extensively evaluate our framework on UltraDomain Agriculture, CS, Legal and Mix, demonstrating its effectiveness and efficiency for graph-based RAG.
\end{itemize}

\section{Related Work}
Graph-based RAG~\cite{peng2024graph} builds graphs characterized by high-level knowledge representations and generate responses through global search. RAPTOR~\cite{sarthi2024raptor} introduces hierarchical tree structures via clustering and summarization, shifting RAG from chunk-based retrieval to structured reasoning. GraphRAG~\cite{edge2024local} further enriches this modeling with diverse graph representations, followed by HiRAG~\cite{huang2025retrieval} and ArchRAG~\cite{wang2025archrag}, which utilize hierarchical community summaries. However, the heavy reliance on frequent LLM calls and extensive summarization makes GraphRAG prohibitively expensive and inefficient for practical deployment.

The subsequent work has been improved in the following two aspects: (1)\textbf{Structural simplification for graph construction.} LightRAG~\cite{guo2024lightrag} integrates lightweight graphs with text indexing and dual-level retrieval for efficient, adaptive knowledge access. KET-RAG~\cite{huang2025ket} combines a sparse knowledge graph skeleton with a text-keyword bipartite graph, enabling multi-granular retrieval without building a full-scale graph. (2)\textbf{Retrieval optimization for global knowledge.} MiniRAG~\cite{fan2025minirag} unifies text and entity indexing into a semantic-aware graph, leveraging lightweight topology-enhanced search for efficient knowledge access. FG-RAG~\cite{hong2025fg} extends entity coverage via context-aware graph retrieval and improves response specificity with query-level fine-grained summarization. PathRAG~\cite{chen2025pathrag} extracts key relational paths and converts them into text prompts, guiding LLMs toward more coherent and context-aware generation.

However, these approaches perform poorly in low-resource scenarios where smaller LLMs are deployed and are rarely explored on incremental knowledge.
TagRAG injects powerful retrieval and reasoning capabilities into an efficiently constructed knowledge graph while supporting incremental knowledge integration.

\section{Task Definition}
TagRAG is dedicated to building a hierarchically explicit knowledge graph under any domain to enable powerful and efficient RAG capabilities.
Given a set of documents $D=\{d_i\}_{i=1}^{\left | D \right | }$, key domain information are extracted to construct an object tag knowledge graph $\mathcal{G}_o=(\mathcal{V}_o, \mathcal{E}_o)$.
In order to form a clear structured knowledge management, given a root domain $\hat{v}$, a hierarchical domain tag system $\mathcal{G}_d=(\mathcal{V}_d, \mathcal{E}_d)$ needs to be established. It contains domain tags $\mathcal{V}_d$ at different levels and hierarchical relationships $\mathcal{E}_d$ between tags.
For knowledge coordination and fusion, the object tags should be linked to the domain tag system and a hierarchical tag knowledge graph is constructed:
\begin{align*}
    \mathcal{G}=(\mathcal{V}_o, \mathcal{E}_o, \mathcal{V}_d, \mathcal{E}_d, \mathcal{E}_{od}),
\end{align*}
where $\mathcal{E}_{od}$ represents the relationships between the object and domain tags.

For retrieval-augmented generation, given a question $q$, the hierarchical tag knowledge graph is searched to generate a global and comprehensive answer $a$:
\begin{align*}
    a \gets \mathcal{F}(q, \mathcal{G}),
\end{align*}
where $\mathcal{F}(\cdot)$ means the graph-based RAG method.

\begin{figure*}[t]
 \centering
 \includegraphics[width=0.98\linewidth]{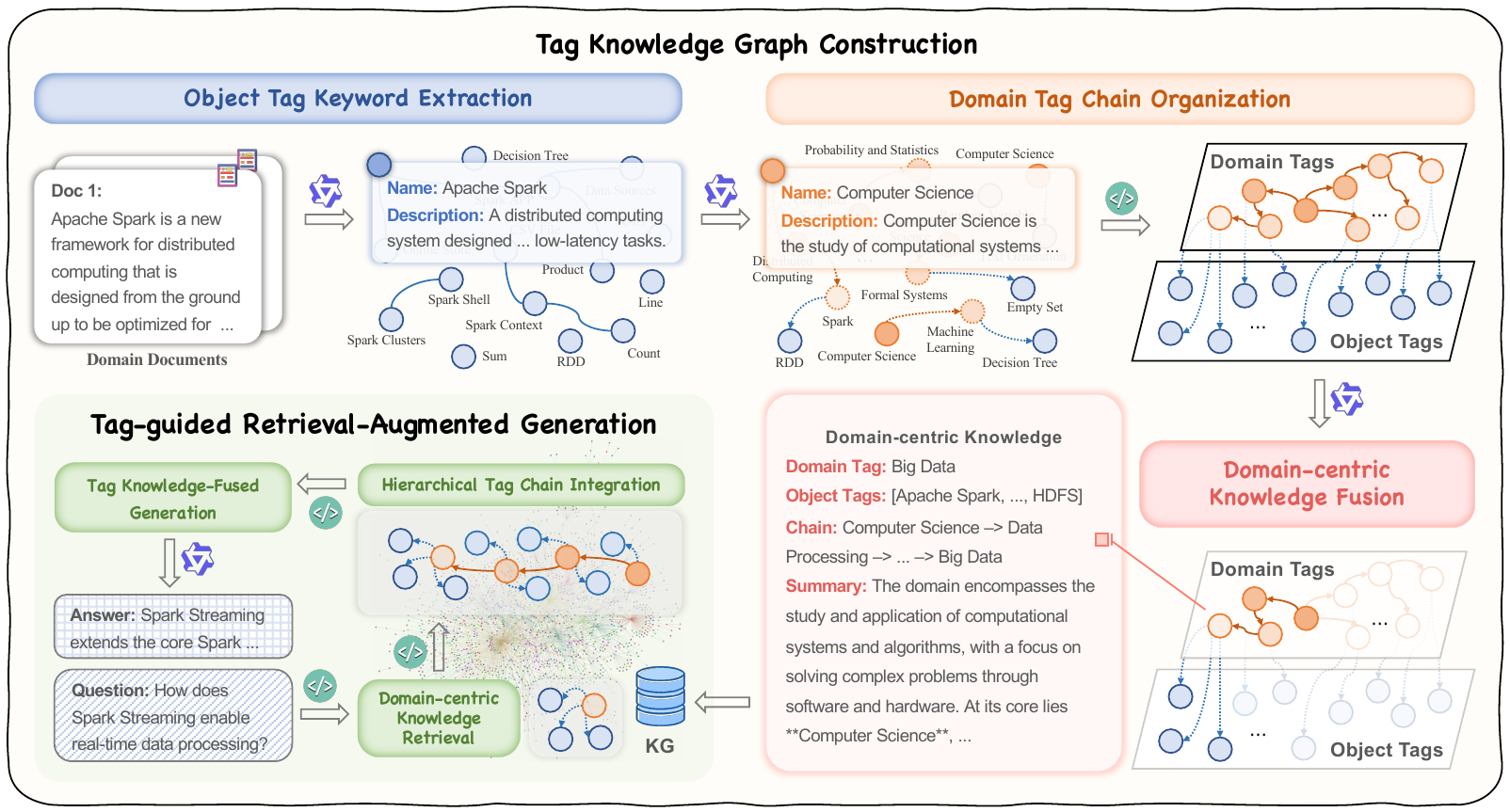}
 \centering
 \caption{The proposed TagRAG framework.}
 \label{fig:framework}
\end{figure*}

\section{TagRAG}
As shown in Figure \ref{fig:framework}, TagRAG consists of two stage: (1) Tag Knowledge Graph Construction and (2) Tag-guided Retrieval-Augmented Generation. 
We extract the keywords from documents as object tags as well as their relationships. Linking them to the pre-defined root domain tag, domain tag chains are organized for knowledge accommodation. 
For information integration, the knowledge from object tags connected to domain tags is fused with summarization.
In the inference stage, domain-centric tags are retrieved for knowledge localization. With the hierarchical tag chains collected, more synthesized information is integrated and can be fused to the global answers.

\subsection{Tag Knowledge Graph Construction}
TagRAG aims to build a knowledge graph with rich domain relationships and efficiently aggregates global knowledge.
We extract object tag keywords to collect expertise, organize domain tag chains to build a hierarchical structured system, and fuse domain-centric knowledge summaries to provide an efficient retrieval platform.

\subsubsection{Object Tag Keyword Extraction}
In order to accurately access the knowledge in the corpus, we need to refine specialized domain documents.
Following the RAG paradigm, the set of documents $D=\{d_i\}_{i=1}^{\left | D \right | }$ are divided into several chunks $T=\{t_i\}_{i=1}^{\left | T \right | }$ with overlaps to reach the size that LLMs can handle and maintain semantic coherence.
Then, we request LLMs to extract domain-specific keywords with their descriptions as object tags:
\begin{align*}
    \mathcal{V}_o, \mathcal{E}_o =  \text{LLM}(\{t_i\}_{i=1}^{\left | T \right | }),
\end{align*}
which contain the most primitive and underlying information in the document.

\subsubsection{Domain Tag Chain Organization}
The extracted object tags are scattered nodes that are difficult to manage hierarchically.
Unlike the bottom-up construction of GraphRAG, we associate them with the predefined root domain tag. This gives the whole construction process a clear sense of direction and satisfies the specialized nature of domain knowledge graphs.
Specifically, given object tags with the predefined root domain tag, we prompt LLMs to generate relationships between them and abstract them into multi-level domain tag chains:
\begin{align*}
    \mathcal{C} = \{c_i\}_{i=1}^{\left | \mathcal{V}_o \right | } =  \text{LLM}(\mathcal{V}_o. \hat{v}),
\end{align*}
Each $c_i$ contains multiple domain tags, each consisting of a domain name and its description. Superior domain tag points to the subordinate domain tag through the relationship of "has subdomain".

% Based on these domain tag chains with explicit pointing relationships, we can merge them into a directed acyclic graph (DAG) to organize hierarchical domain knowledge.
These domain tag chains, featuring explicit pointing relationships, inherently carry the hierarchical knowledge logic from general categories to specific subfields. To consolidate this logic into a structured framework that avoids redundant associations and cyclic dependencies, we merge the chains into a directed acyclic graph (DAG) that clearly embodies the layered hierarchy of domain knowledge while ensuring intuitive traversal between different levels of tags.
Starting at the head of the chain, each node is mounted at its parent's corresponding position in the DAG to ensure the integrity of the chain and the transitivity of the DAG.
% Considering incremental insertion, for nodes with the same name, merge their descriptions, otherwise insert new nodes.
The detailed process of domain tag chain organization can be seen in Algorithm~\ref{algo:chain_organization}.

\begin{algorithm}[h!]
    \KwIn{The list of domain tag chain $\mathcal{C}$, the root domain tag $\hat{v}$ and the existing tag knowledge graph $\mathcal{G}$.}
    \KwOut{$\mathcal{\hat{G}}$.}
    
    $\text{DAG} \leftarrow \{\hat{v}\}$\;
    \ForEach{$c_i \in \mathcal{C}$}{
        \For{$j$ from $2$ to $\left | c_i \right | - 1$}{
            $p_{i,j} \leftarrow \text{get\_node}(\text{DAG}, c_{i,j-1})$\;
            $n_{i,j} \leftarrow \text{get\_node}(\text{DAG}, c_{i,j})$\;
            \If{$p_{i,j}$ is not None}{
                \If{$n_{i,j}$ is not None}{
                    \If{$n_{i,j} \notin p_{i,j}.\text{children}$}{
                        $p_{i,j}.\text{add\_child}(n_{i,j})$\;
                    }
                }
                \Else{
                    $n_{new} \leftarrow \text{new\_node}(c_{i,j})$\;
                    $p_{i,j}.\text{add\_child}(n_{new})$\;
                }
            }
        }
    }
    $\mathcal{\hat{G}} \leftarrow \text{merge\_graph}(\mathcal{G}, \text{DAG})$\;
    \Return{$\mathcal{\hat{G}}$}
    \caption{Domain Tag Chain Organization}
    \label{algo:chain_organization}
\end{algorithm}

\subsubsection{Domain-centric Knowledge Fusion}
In order to efficiently retrieve domain information, we deploy knowledge fusion in the graph construction phase in advance, which aggregates all highly relevant information into domain tags.
Specifically, there are two such categories of information:

\begin{itemize}
    \item \textbf{Domain information on chains}: there are multiple domain nodes on domain tag chains that contain rich high-level domain information. They are natural providers of comprehensive field vision.
    \item \textbf{Specialized knowledge in objects}: the extracted object tags contain the expertise in the original documents. They can provide accurate grounding knowledge.
\end{itemize}

Combining these two aspects of data, we fused them into the domain-centric knowledge summary:
\begin{align*}
    s = \text{LLM}(\text{Chain}(v_d), \text{Nei}(v_d)),
\end{align*}
where $v_d \in V_d$ denotes each domain tag, $\text{Chain}(\cdot)$ indicates the domain tag chain to which $v_d$ belongs, and $\text{Nei}(\cdot)$ represents the object tags linked to $v_d$. 

Embedded in the summary, each domain tag is accompanied by global knowledge related to itself.
After vectorizing these summaries, we obtain a diverse and global domain-centric knowledge retrieval library:
\begin{align*}
    \mathcal{K} \leftarrow \{v_i, s_i, \text{Emb}(s_i)\}_{i=1}^{\left | V_d \right |},
\end{align*}
where $\text{Emb}(\cdot)$ is the embedding function, $v_i$ and $s_i$ denote a domain tag and its summary,   respectively.

\subsubsection{Knowledge Incremental Insertion}
The TagRAG framework has a natural advantage in knowledge incremental insertion.
Instead of dividing communities from scratch, TagRAG can directly embed newly constructed domain tag chains into the existing knowledge graph.
The incremental process involves two components:

\begin{itemize}
    \item \textbf{Tag increment}: For the new extracted object tags or generated domain tags, TagRAG appends new descriptions to the existing ones with the same name.
    \item \textbf{Knowledge increment}: For the new fused domain-centric knowledge, TagRAG re-summarizes old and new summaries of domain tags with the same name to generate new perceptions.
\end{itemize}

Compared to the time-consuming full reconstruction of GraphRAG, TagRAG’s incremental insertion mechanism is significantly more efficient.
% Compared to the time-consuming full reconstruction required by GraphRAG, TagRAG’s incremental insertion mechanism significantly improves efficiency, enabling real-time updates with low resource consumption.

\subsection{Tag-guided Retrieval-Augmented Generation}
Following the global vision strategy, TagRAG uses tag-guided graph retrieval-augmented generation.
As the existence of domain tags that carry global knowledge, TagRAG is more efficient than other summarization-based and walk-based methods.

\subsubsection{Domain-centric Knowledge Retrieval}
Based on the domain-centric knowledge retrieval library, we can search domain knowledge related to the question:
\begin{align*}
    (V_t^{'}, S_t^{'}) = \text{Ret}_{\text{tag}}(q, \mathcal{K}),
\end{align*}
where $V_t^{'}$ indicates the related domain tags, $S_t^{'}$ represents the related summaries, and $\text{Ret}_{\text{tag}}(\cdot)$ is the retrieval function with cosine similarity.

\subsubsection{Hierarchical Tag Chain Integration}
Thanks to knowledge fusion, domain-centric knowledge already has global properties. 
However, with the advantage of hierarchical knowledge graph, we are able to further extract higher-level global knowledge:
\begin{align*}
    (V_c^{'}, S_c^{'}) = \text{Ret}_{\text{chain}}(V_t^{'}),
\end{align*}
where $\text{Ret}_{\text{chain}}(\cdot)$ extracts the corresponding chains, $V_c^{'}$ indicates the domain tags on the chain corresponding to $V_t^{'}$, and $S_c^{'}$ represents the summaries implied by $V_c^{'}$.

\subsubsection{Tag Knowledge-Fused Generation}
With the retrieved summaries, global information is achieved for response generation.
Constrained by the input length of the LLMs, we prioritize putting in the related domain tag summaries $S_t^{'}$, and then adding the related chain summaries $S_c^{'}$ until the upper limit is reached.
Given the question and the domain knowledge summaries, the answer is generated by the LLM:
\begin{align*}
    a = \text{LLM}(q, S_t^{'}, S_c^{'}).
\end{align*}
\vspace{-20pt}

\subsection{Analysis of Retrieval Complexity}
GraphRAG's inefficiency stems from its multi-stage pipeline, requiring extensive LLM calls for entity extraction and community detection (e.g., Leiden algorithm), leading to high computational costs and frequent full graph reconstructions for dynamic data. In contrast, TagRAG's domain tag chains enable hierarchical information aggregation during graph building, replacing iterative community partitioning with linear chain processing. During inference, TagRAG utilizes vector matching and tag chain linking, avoiding GraphRAG's costly full graph traversal. Both stages demonstrate TagRAG's superior efficiency over GraphRAG.

\section{Experimental Setup}

% \subsection{Datasets}
% To demonstrate the high applicability of TagRAG, following standard evaluation on Graph-based RAG methods~\cite{edge2024local,guo2024lightrag,chen2025pathrag}, we chose four corpus from the comprehensive UltraDomain~\cite{qian2025memorag} benchmark, Agriculture, CS, Legal and Mix. The size of each dataset ranges from 600,000 to 5,000,000 tokens.
% Following LightRAG, we used GPT-4o-mini to generate 125 global questions for each dataset, covering different domains and different tasks.
% The details of the datasets can be seen in Table~\ref{tab:dataset details}.
% The question generation prompt can be seen in Appendix.

% \begin{table}[h]
% \centering
% \small
% \begin{tabular}{lllll}
% \toprule
%  & Agri & CS & Legal & Mix \\ \midrule
% Docs & 12 & 10 & 94 & 61 \\
% Chunks & 1756 & 1858 & 4294 & 579 \\
% Tokens & 2,017,886 & 2,306,535 & 5,081,069 & 619,009 \\
% Size & 8.56MB & 8.51MB & 21.24MB & 2.54MB \\ 
% \bottomrule
% \end{tabular}
% \caption{Dataset Details.}
% \label{tab:dataset details}
% \end{table}

\subsection{Datasets and Baselines}
To demonstrate the high applicability of TagRAG, following standard evaluation on Graph-based RAG methods~\cite{edge2024local,guo2024lightrag,chen2025pathrag}, we chose four corpus from the comprehensive UltraDomain~\cite{qian2025memorag} benchmark, Agriculture, CS, Legal and Mix.
% The size of each dataset ranges from 600,000 to 5,000,000 tokens.
Following LightRAG, we used GPT-4o-mini to generate 125 global questions for each dataset, covering different domains and different tasks.
% The details of the datasets can be seen in Table~\ref{tab:dataset details}.
The dataset details and question generation prompt can be seen in Appendix~\ref{app:dataset} and \ref{app:prompts}.

% \begin{table}[h]
% \centering
% \small
% \begin{tabular}{lllll}
% \toprule
%  & Agri & CS & Legal & Mix \\ \midrule
% Docs & 12 & 10 & 94 & 61 \\
% Chunks & 1756 & 1858 & 4294 & 579 \\
% Tokens & 2,017,886 & 2,306,535 & 5,081,069 & 619,009 \\
% Size & 8.56MB & 8.51MB & 21.24MB & 2.54MB \\ 
% \bottomrule
% \end{tabular}
% \caption{Dataset Details.}
% \label{tab:dataset details}
% \end{table}

To validate the performance and efficiency of global question answering, we compare TagRAG with two types of baselines. \textbf{(1) Zero-shot LLM Generation:} We call Qwen3-4B~\cite{qwen3technicalreport}, Qwen3-30B-A3B~\cite{qwen3technicalreport} and Llama-3.3-70B-Instruct~\cite{grattafiori2024llama3herdmodels} to directly answer the questions. \textbf{(2) Retrieval-Augmented Generation:} NaiveRAG~\cite{lewis2020retrieval} focuses on local context through dynamic document retrieval. GraphRAG~\cite{edge2024local} utilizes entity graphs and community summaries for global knowledge synthesis. LightRAG~\cite{guo2024lightrag} is a lightweight baseline that integrates graph structures with dual-level text indexing for fast and real-time updates. MiniRAG~\cite{fan2025minirag} achieves high efficiency on small language models via semantic-aware graph indexing and lightweight topology-based retrieval. Detailed descriptions are shown in Appendix~\ref{app:baseline}.

\subsection{Evaluation Metrics}
Four evaluation metrics were used to assess the performance of the comparison methods:
\textbf{Comprehensiveness}: How much detail does the answer provide to cover all aspects and details of the question?
\textbf{Diversity}: How varied and rich is the answer in providing different perspectives and insights on the question?
\textbf{Empowerment}: How well does the answer help the reader understand and make informed judgments about the topic?
\textbf{Overall}: Which answer is better overall?

% \begin{itemize}
%     \item \textbf{Comprehensiveness}: How much detail does the answer provide to cover all aspects and details of the question?
%     \item \textbf{Diversity}: How varied and rich is the answer in providing different perspectives and insights on the question?
%     \item \textbf{Empowerment}: How well does the answer help the reader understand and make informed judgments about the topic?
%     \item \textbf{Overall}: Which answer is better overall?
% \end{itemize}

We utilize the powerful model GPT-4o-mini, gemini-2.5-pro and claude-sonnet-4.5-20250929 to determine the winner for each of the two comparison methods based on the above metrics. For each dataset, we also exchange the order of the results of the two comparison methods to avoid position bias. Finally, we report the average of 3*2=6 evaluation results.
% We utilize the powerful model GPT-4o-mini to determine the winner for each of the two comparison methods based on the above metrics at once and calculate the winning rate on each dataset.
The evaluation prompt can be seen in Appendix~\ref{app:prompts}.

\subsection{Implementation Details}
We use Qwen3-4B~\cite{qwen3technicalreport} as the backbone to conduct the experiments without thinking. bge-large-en-v1.5~\cite{bge_embedding} is employed to embed questions and documents. 
The chunk size and overlap size are 1200 and 100, respectively. The vector database used in this work is nano-vectordb~\footnote{\url{https://github.com/gusye1234/nano-vectordb}}. 
The top-k number for Domain-centric Knowledge Retrieval is 3. 
% All experiments are run on a Linux server with Intel(R) Xeon(R) Gold 6330 CPU @ 2.00GHz, 251G RAM, NVIDIA RTX A6000, CentOS Linux 7, CUDA 12.4 and python 3.10.16.
% All experiments are completed on a Linux server with NVIDIA RTX A6000.
More detailed experimental settings can be found in Appendix~\ref{app:experimental details}.

\begin{table}[t!]
\centering
\small
% \begin{tabular}{lrrrrr}
\begin{tabularx}{0.48\textwidth}{lXXXXX}
\toprule
& Agri & CS & Legal & Mix & Avg \\ \midrule
\rowcolor{gray!20} 
\multicolumn{6}{c}{\textbf{Zero-shot LLM Generation}}\\ \midrule
\multicolumn{6}{c}{\textit{v.s. Qwen3-4B}}\\ \midrule
Comprehensiveness & 85.9 & 69.6 & 70.3 & 58.8 & 71.2 \\
Diversity         & 78.9 & 64.4 & 62.0 & 53.9 & 64.8 \\
Empowerment       & 80.7 & 57.5 & 48.9 & 45.7 & 58.2 \\
Overall           & 83.2 & 59.3 & 53.7 & 49.3 & 61.4 \\ \midrule

\multicolumn{6}{c}{\textit{v.s. Qwen3-30B-A3B}}\\ \midrule
Comprehensiveness & 84.8 & 68.8 & 72.7 & 59.6 & 71.5 \\
Diversity         & 79.5 & 65.1 & 62.1 & 59.6 & 66.6 \\
Empowerment       & 76.4 & 53.2 & 45.6 & 40.4 & 53.9 \\
Overall           & 80.3 & 55.5 & 51.6 & 44.7 & 58.0 \\ \midrule

\multicolumn{6}{c}{\textit{v.s. Llama-3.3-70B-Instruct}}\\ \midrule
Comprehensiveness & 68.1 & 51.5 & 49.7 & 27.7 & 49.2 \\
Diversity         & 67.5 & 58.5 & 43.1 & 35.9 & 51.2 \\
Empowerment       & 56.9 & 43.1 & 27.2 & 22.8 & 37.5 \\
Overall           & 65.9 & 49.3 & 39.7 & 27.3 & 45.5 \\ \midrule

\rowcolor{gray!20} 
\multicolumn{6}{c}{\textbf{Retrieval-Augmented Generation}}\\ \midrule
\multicolumn{6}{c}{\textit{v.s. NaiveRAG}}\\ \midrule
Comprehensiveness & 93.9 & 94.0 & 79.5 & 90.0 & 89.3 \\
Diversity         & 90.4 & 95.5 & 95.2 & 95.2 & 94.1 \\
Empowerment       & 91.2 & 88.3 & 74.9 & 88.1 & 85.6 \\
Overall           & 93.5 & 89.7 & 72.3 & 87.9 & 85.8 \\ \midrule
\multicolumn{6}{c}{\textit{v.s. GraphRAG}}\\ \midrule
Comprehensiveness & 89.3 & 85.3 & 60.7 & 67.1 & 75.6 \\
Diversity         & 88.5 & 85.5 & 72.0 & 74.9 & 80.2 \\
Empowerment       & 90.8 & 81.6 & 63.2 & 67.9 & 75.9 \\
Overall           & 90.9 & 81.9 & 62.3 & 67.7 & 75.7 \\ \midrule
\multicolumn{6}{c}{\textit{v.s. LightRAG}}\\ \midrule
Comprehensiveness & 94.4 & 85.1 & 80.3 & 88.1 & 87.0 \\
Diversity         & 94.9 & 88.1 & 88.5 & 92.3 & 91.0 \\
Empowerment       & 95.5 & 85.7 & 82.0 & 91.6 & 88.7 \\
Overall           & 94.4 & 84.0 & 79.9 & 89.7 & 87.0 \\ \midrule
\multicolumn{6}{c}{\textit{v.s. MiniRAG}}\\ \midrule
Comprehensiveness & 76.0 & 55.2 & 56.8 & 64.1 & 63.0 \\
Diversity         & 80.7 & 64.7 & 70.4 & 77.1 & 73.2 \\
Empowerment       & 79.1 & 59.7 & 59.5 & 67.3 & 66.4 \\
Overall           & 78.3 & 58.4 & 57.5 & 65.3 & 64.9 \\
\bottomrule
\end{tabularx}
% \end{tabular}
\caption{Main results: winning rates (\%) of TagRAG v.s. baselines with Qwen3-4B across four datasets.}
\label{tab:main results}
\end{table}

\begin{figure*}[t]
    \centering
    \begin{tabular}{cc}
        \includegraphics[width=0.48\linewidth]{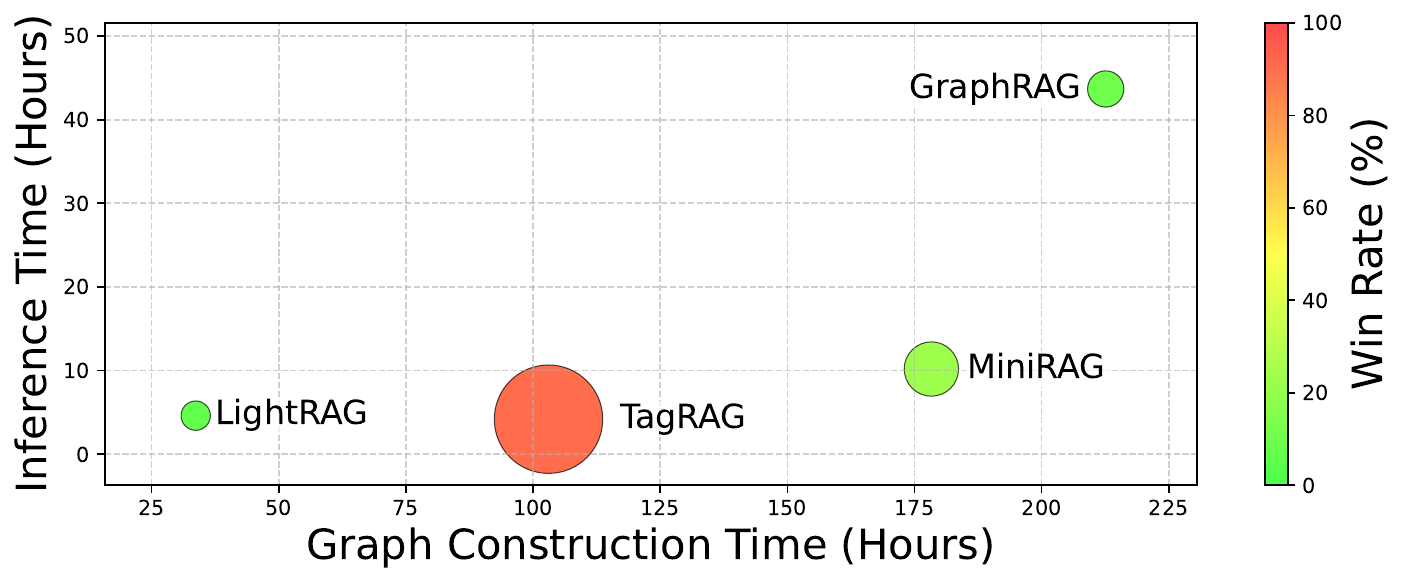} &
        \includegraphics[width=0.48\linewidth]{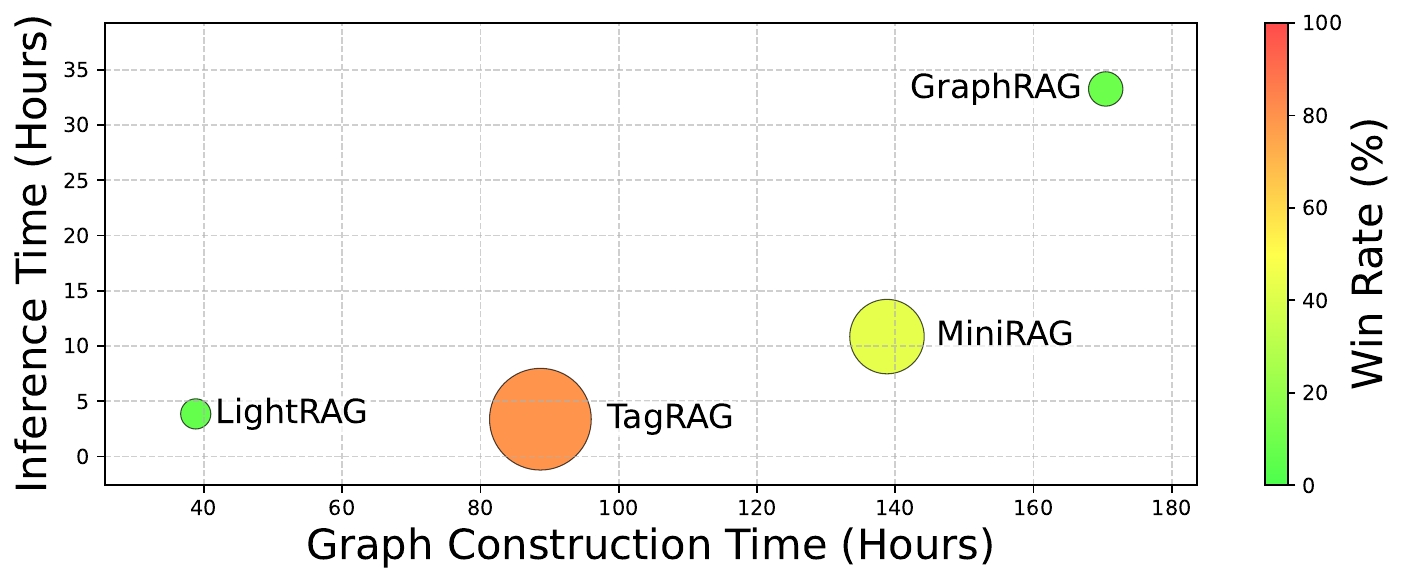} \\
        (a) On UltraDomain Agriculture & (b) On UltraDomain CS \\
        \includegraphics[width=0.48\linewidth]{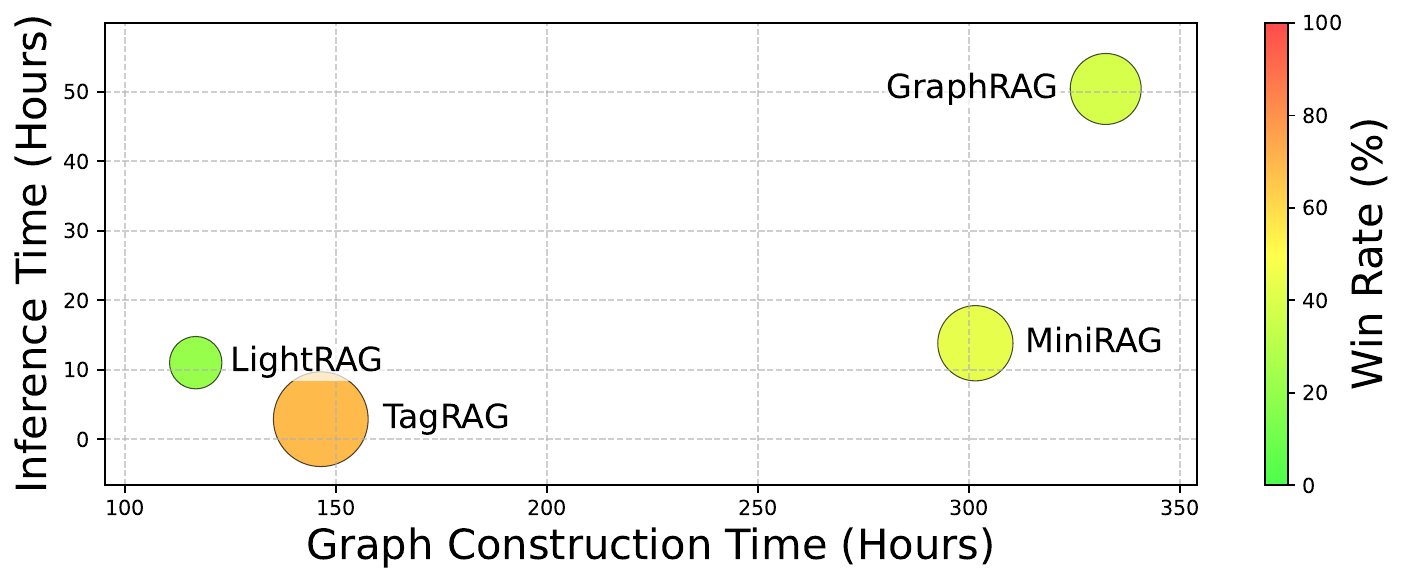} &
        \includegraphics[width=0.48\linewidth]{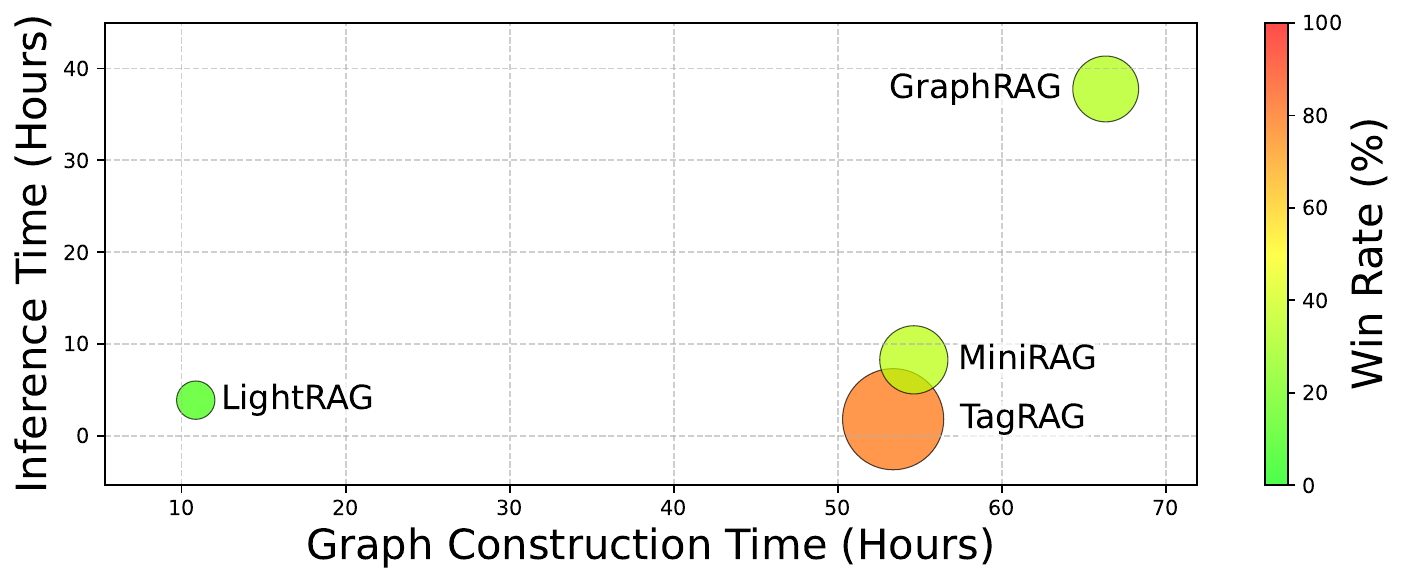} \\
        (c) On UltraDomain Legal & (d) On UltraDomain Mix \\
    \end{tabular}
    \caption{Performance-efficiency analysis: comparative winning rates, graph construction time and inference time results of TagRAG and baselines across four datasets. The larger the bubble and the closer to the lower left corner, the better the method.}
    \label{fig:performance-efficiency analysis}
\end{figure*}

\begin{figure*}[t]
    \centering
    \begin{tabular}{cc}
        \includegraphics[width=0.48\linewidth]{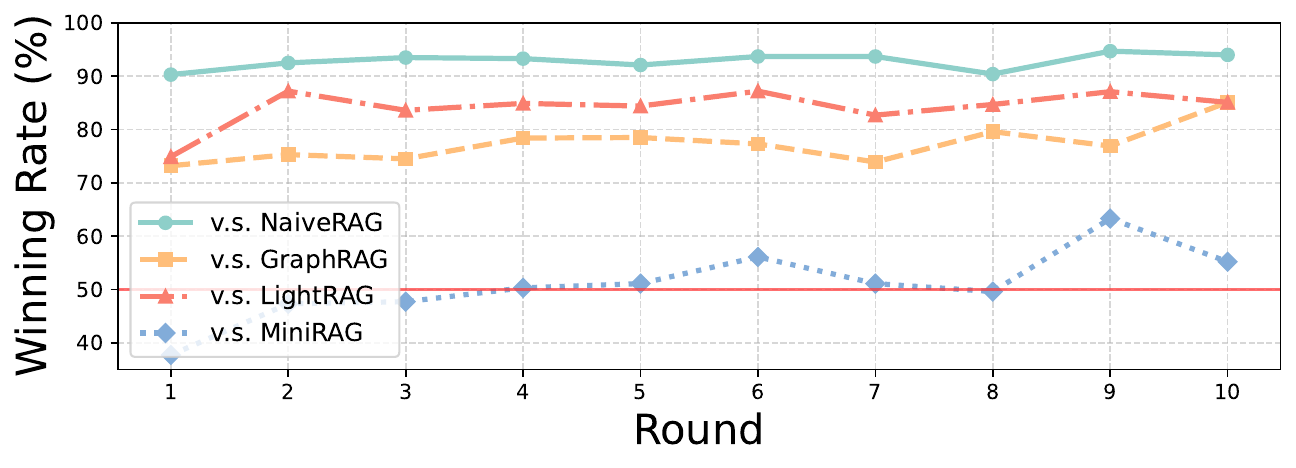} &
        \includegraphics[width=0.48\linewidth]{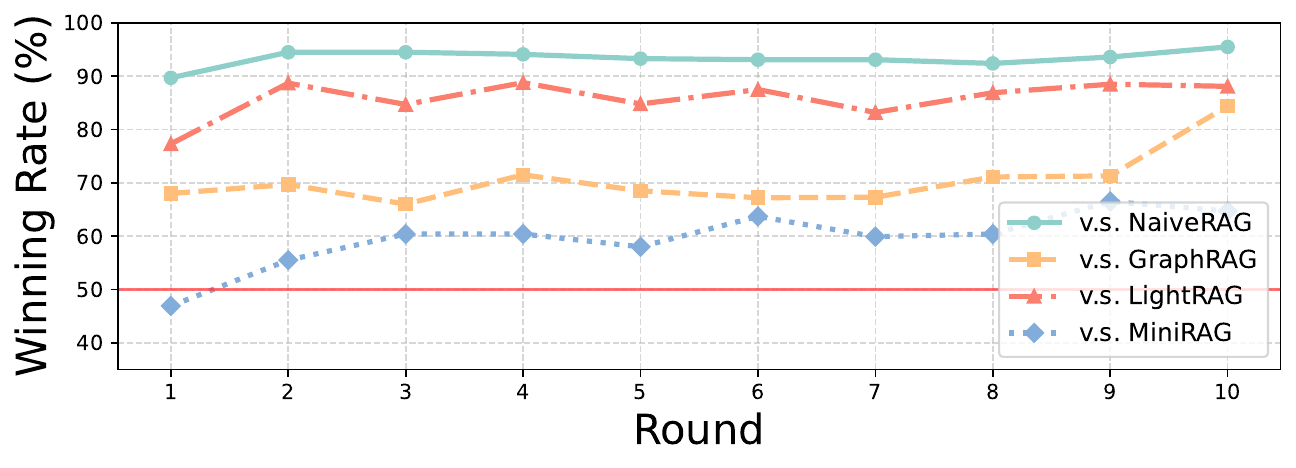} \\
        (a) Comprehensiveness & (b) Diversity \\
        \includegraphics[width=0.48\linewidth]{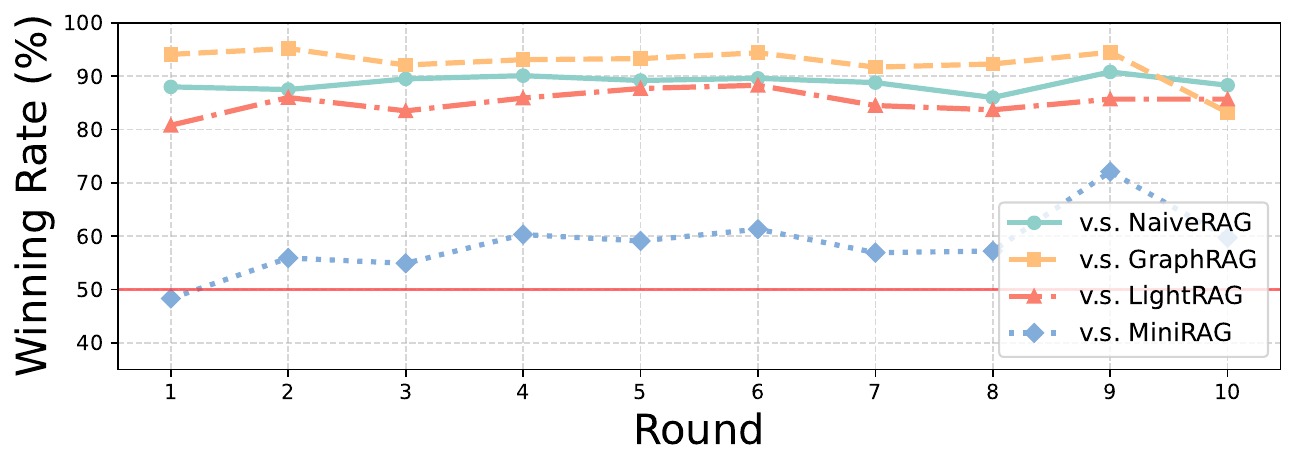} &
        \includegraphics[width=0.48\linewidth]{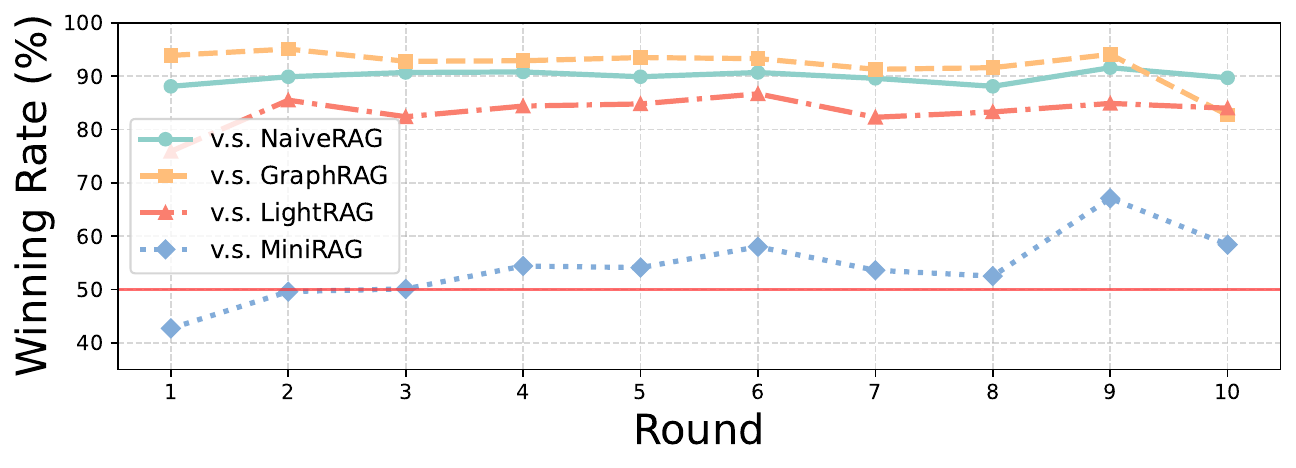} \\
        (c) Empowerment & (d) Overall \\
    \end{tabular}
    \caption{Incremental analysis: winning rates (\%) of TagRAG v.s. baselines with Qwen3-4B across four datasets in 10 rounds. The red horizontal line represents equilibrium and values above it indicate TagRAG's advantage.}
    \label{fig:incremental analysis}
\end{figure*}

\section{Results and Analysis}

% \subsection{Main Results}
% Table 1 evaluates the winning rate of TagRAG against NaiveRAG, GraphRAG, LightRAG and MiniRAG on four datasets, UltraDomain Agriculture, CS, Legal and Mix.
% We can draw the following conclusions:

% \twbadd{
% \begin{itemize}
%     \item \textbf{TagRAG absolutely lead in performance.} Compared with NaiveRAG, GraphRAG and LightRAG, TagRAG achieves an average winning percentage of 82.85\% percent. MiniRAG, the best performer in Baselines, retains just a 35.125\% winning rate in average against TagRAG.
%     \item \textbf{TagRAG excels at Diversity.} With the global nature of Domain Knowledge Integration, TagRAG has access to diverse information for generation. Even when faced with GraphRAG traversing all communities for large amounts of information, TagRAG still maintains clear advantages.
% \end{itemize}
% }q
\subsection{Main Results} 
Table 1 evaluates the winning rate of TagRAG against NaiveRAG, GraphRAG, LightRAG, and MiniRAG on four datasets: UltraDomain Agriculture, CS, Legal, and Mix. We can draw the following conclusions: (1)\textbf{TagRAG leads in performance}, achieving an average winning percentage of 82.85\% compared with NaiveRAG, GraphRAG, and LightRAG. Even against MiniRAG, the strongest baseline, TagRAG maintains a dominant average winning rate, with MiniRAG retaining only 35.125\%. (2)\textbf{TagRAG excels at Diversity}, leveraging its domain knowledge integration to access broader information for generation. Even when compared to GraphRAG, which traverses all communities, TagRAG maintains a clear advantage, suggesting that fusing knowledge via domain chains is superior to community-summary or path-discovery approaches. \textbf{(3)TagRAG expands the model capability}, because it defeats the LLM with 7 times the parameters by virtue of the fused domain chain information. Even facing the 70B model, TagRAG can compete with it. This demonstrates that TagRAG does not rely solely on the inherent capabilities of the language model, and that the tag chain mechanism effectively broadens the boundaries of global reasoning.

\begin{table}[h]
\centering
\small
% \begin{tabular}{lrrrr}
\begin{tabularx}{0.48\textwidth}{lXXXX}
\toprule
& Agri& CS& Legal& Mix\\ \midrule
\multicolumn{5}{c}{\textit{v.s. w/o chain}}\\   \midrule
Comprehensiveness & 87.1& 90.5& 85.2& 78.9\\
% Directness & 87.2 & 12.8 & 84.0 & 16.0 & 92.0 & 8.0 & 80.8 & 19.2  \\
Diversity & 84.7& 87.3& 85.6& 76.3\\
Empowerment & 86.8& 84.5& 77.7& 72.8\\
Overall & 87.2& 87.5& 80.1& 75.2\\  \midrule

\multicolumn{5}{c}{\textit{v.s. w/o fusion}}\\   \midrule
Comprehensiveness & 97.3& 95.5& 95.5& 85.9\\
% Directness & 83.2 & 16.8 & 78.4 & 21.6 & 88.8 & 11.2 & 92.0 & 8.0   \\
Diversity & 96.4& 94.1& 94.5& 89.1\\
Empowerment & 96.9& 87.1& 85.1& 76.9\\
Overall & 96.9& 89.7& 88.0& 78.4\\
\bottomrule
\end{tabularx}
% \end{tabular}
\caption{Ablation study: winning rates (\%) of TagRAG v.s. w/o chain and w/o fusion with Qwen3-4B across four datasets.}
\label{tab:ablation study}
\end{table}

\subsection{Performance-efficiency Analysis}
In Figure \ref{fig:performance-efficiency analysis}, we show the performance, graph construction time and inference time of different methods. TagRAG, which presents the largest bubbles, embodies the strongest performance, and possesses low graph construction time and inference time.
GraphRAG not only exhibits weak performance, but also consumes extremely high construction and inference costs.
MiniRAG has the next strongest performance, but its graph construction time is far behind TagRAG, due to its ultimate entity and relationship extraction.
Despite the excellent time overhead with lightweight extraction, the performance of LightRAG is the worst, subject to the gap of high and low level semantics.
Taken together, thanks to the advanced domain chain indexing and knowledge integration, TagRAG has achieved the strongest performance-efficiency results.

\begin{table*}[t]
\centering
\small
\begin{tabular}{lrllrllrllrll}
\toprule
& \multicolumn{3}{c}{\textit{v.s. NaiveRAG}}& \multicolumn{3}{c}{\textit{v.s. GraphRAG}}& \multicolumn{3}{c}{\textit{v.s. LightRAG}}& \multicolumn{3}{c}{\textit{v.s. MiniRAG}}\\ \midrule
&\textit{large}&\textit{base}&\textit{small}&  \textit{large}& \textit{base}&\textit{small}&   \textit{large}&\textit{base}&\textit{small}& \textit{large}& \textit{base}&\textit{small}\\ \midrule
Comprehensiveness & 94.0& 93.7&93.1& 85.3& 86.0&86.1& 85.1&81.5&82.3& 55.2& 57.5&59.9\\
Diversity & 95.5& 95.6&96.3& 85.5& 85.6&87.2& 88.1&87.3&86.8& 64.7& 69.2&66.0\\
Empowerment & 88.3& 90.9&90.5& 81.6& 84.4&85.2& 85.7&84.9&85.7& 59.7& 62.8&64.5\\
Overall & 89.7& 89.9&91.1& 81.9& 82.8&85.1& 84.0&82.5&83.3& 58.4& 60.7&62.8\\
\bottomrule
\end{tabular}
\caption{Retriever adaption analysis: winning rates (\%) of TagRAG v.s. baselines with Qwen3-4B on CS, equiped with different retrievers of bge-large-en-v1.5, bge-base-en-v1.5 and bge-small-en-v1.5.}
\label{tab:retriever adaption analysis}
\end{table*}

\subsection{Ablation Study}
We conduct ablation experiments on the four datasets by comparing TagRAG with TagRAG w/o chain and w/o fusion. 
TagRAG w/o chain simply adds the retrieved fused domain-centric knowledge to the context without involving the information in the associated domain chain.
And TagRAG w/o fusion not only discards the knowledge of the domain chain, but also cuts off the object tag information connected to the retrieved domain tag. It relies entirely on the description of the domain tags to answer the question.
It can be clearly found that (1) the absence of these two components makes the answers generated by the model far inferior to the full TagRAG and (2) retaining fused domain-centric knowledge is better than simply using the description of domain tags for generation.
% Unexpectedly, integrating more relevant knowledge leads to more comprehensive and integrated answers.

\subsection{Incremental Analysis}
To validate that our proposed method can accommodate incremental scenarios, we incrementally construct a knowledge graph from 10 documents in the UltraDomain CS dataset. After each round of increment, we generate diverse questions involving all current documents to test the global reasoning capabilities of different methods.
In Figure~\ref{fig:incremental analysis}, Tag RAG has a steady lead over Naive RAG, Graph RAG and Light RAG, with average winning rates reaching over 80\%.
Even though there is a close tie with TagRAG in the first round, MiniRAG loses as the documents increase.
This proves that TagRAG has a stable global inference capability in incremental scenarios with multiple rounds, as the mechanism of tag chain fusion naturally has the advantage of infinite scaling.

\subsection{Retriever adaption analysis}

%To demonstrate that our method does not rely on a powerful searcher, we performed a searcher adaptability analysis.
To demonstrate that our performance does not solely depend on the effect of a powerful searcher, we perform a retriever adaption analysis.
As shown in Table~\ref{tab:retriever adaption analysis}, retrievers of different sizes are applied to the comparison.
Even with a lower performance retriever, TagRAG still has a stable advantage in all metrics. Even in comparison with GraphRAG, the poorer retriever excites the potential of TagRAG.
This suggests that domain clustering in the form of tag chains based on DAG weakens the need for precise entity retrieval and enhances the representation of global knowledge.

\subsection{Cross-domain incremental analysis}
% Although it has been verified on the Mix dataset, we add a document from the UltraDomain CS dataset to the knowledge graph already constructed from the Mix dataset, to demonstrate the cross-domain capabilities of the comparative methods.
% After integrating the new data, we let the comparative methods still answer the questions generated for the Mix dataset to determine whether they can fully utilize the incrementally constructed knowledge to make the responses better than before.
% In Table \ref{tab:cross-domain incremental analysis}, TagRAG substantially outperforms the other three methods in terms of performance and achieves suboptimal results in the time dimension.
% LightRAG, despite its time advantage, is limited by its polarized knowledge representation, which makes it difficult to merge new knowledge well into the already existing framework, leading to the unsatisfactory incremental performance.
% GraphRAG unquestionably consumes the most time due to its large number of LLM calls as well as community summarizations, and it struggles to incorporate knowledge from completely new domains into the established communities, leading to its poor performance.
To demonstrate cross-domain capabilities, we add a document from the UltraDomain CS dataset to the Mix-built knowledge graph. After integration, we evaluate whether comparative methods can leverage this incremental knowledge to improve responses to Mix-generated questions.
As shown in Table \ref{tab:cross-domain incremental analysis}, TagRAG substantially outperforms the other methods in performance while achieving suboptimal time results.
LightRAG, despite its time advantage, is limited by polarized knowledge representation, hindering effective merging into the existing framework and leading to unsatisfactory incremental performance.
GraphRAG consumes the most time due to numerous LLM calls and community summarizations, struggling to incorporate entirely new domains into established communities, resulting in poor performance.

\begin{table*}[t]
\centering
\small
\begin{tabular}{lrrrrrr}
\toprule
& Comprehensiveness & Diversity & Empowerment & Overall & Time-C & Time-I \\ \midrule
GraphRAG & 41.7& 42.8& 43.2& 44.0& 30.47 & 36.81 \\
LightRAG & 53.5& 54.5& 52.9& 52.9& 2.28 & 4.01 \\
MiniRAG & 53.9& 53.2& 52.9& 54.1& 9.83 & 8.80 \\
TagRAG & 56.1& 56.1& 56.8& 58.0& 6.37 & 2.47 \\
\bottomrule
\end{tabular}
\caption{Cross-domain incremental analysis: winning rates(\%) of results after v.s. before upserting. Time-C (hours) and Time-I (hours) denote incremental graph construction time and inference time, respectively.}
\label{tab:cross-domain incremental analysis}
\end{table*}

\subsection{Visualization of Knowledge Graph Construction}

\begin{figure}[h]
    \centering
    \begin{tabular}{cc}
        \includegraphics[width=0.45\linewidth]{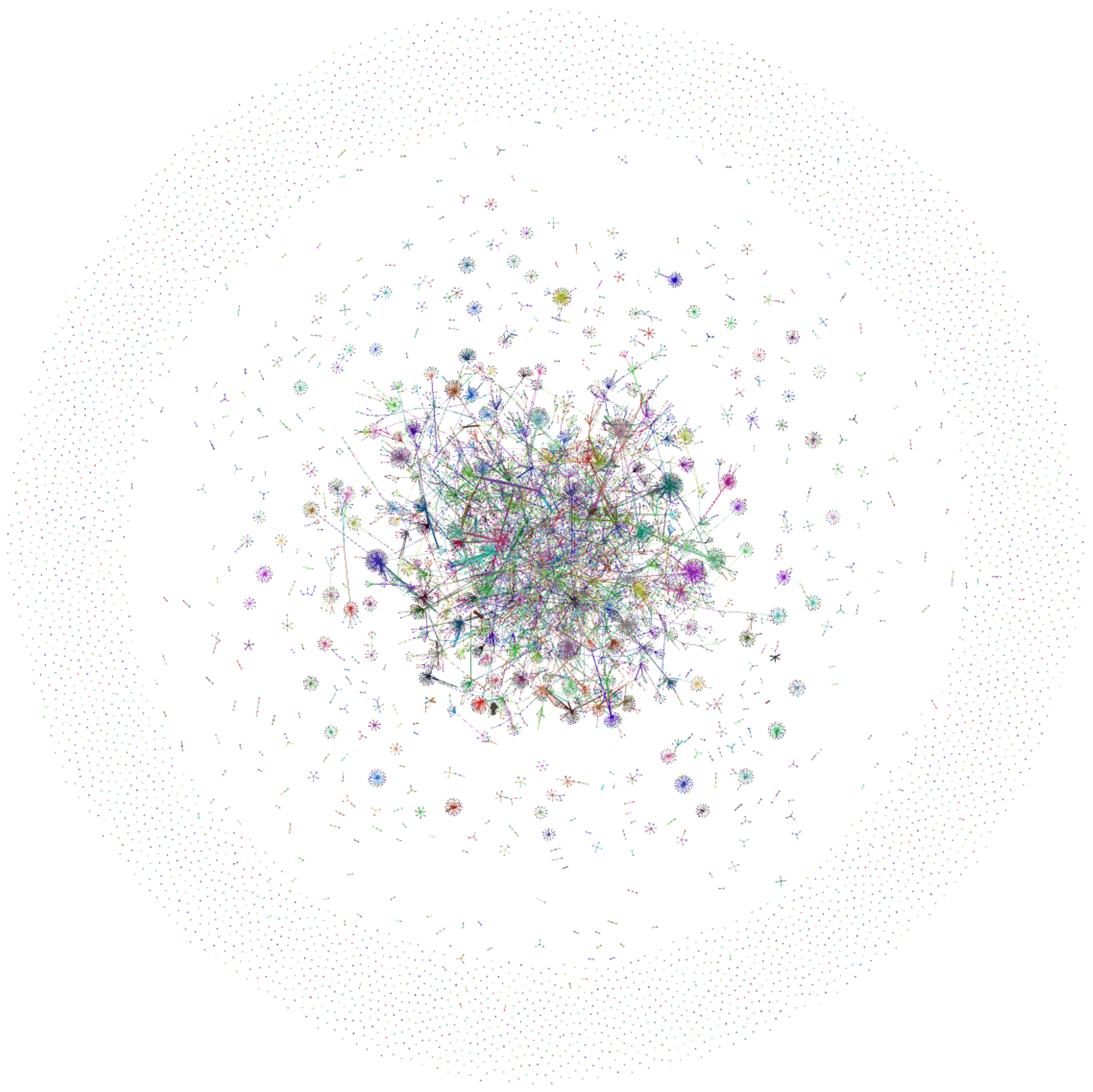} &
        \includegraphics[width=0.45\linewidth]{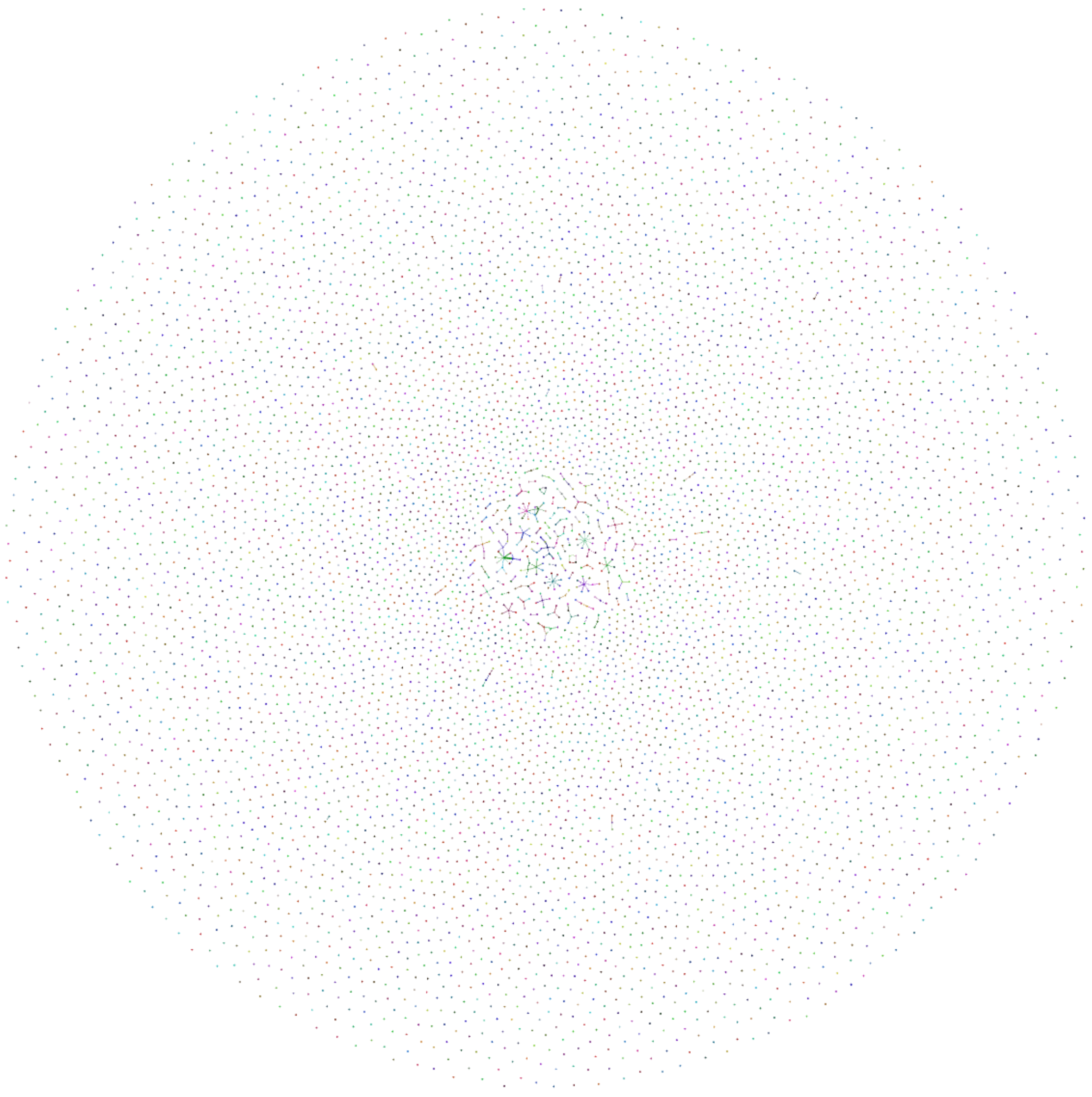} \\
        (a) GraphRAG & (b) LightRAG \\
        \includegraphics[width=0.45\linewidth]{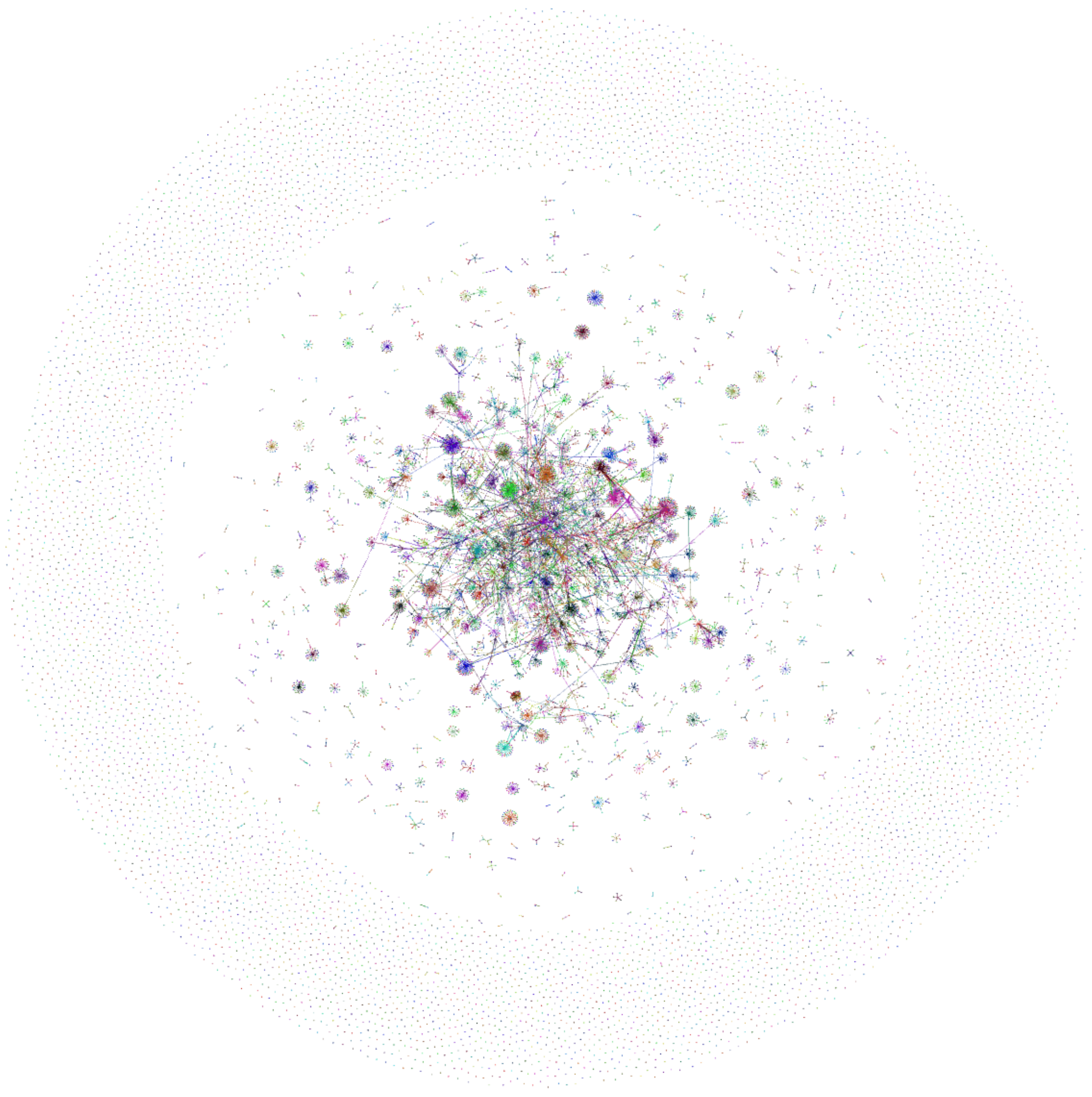} &
        \includegraphics[width=0.45\linewidth]{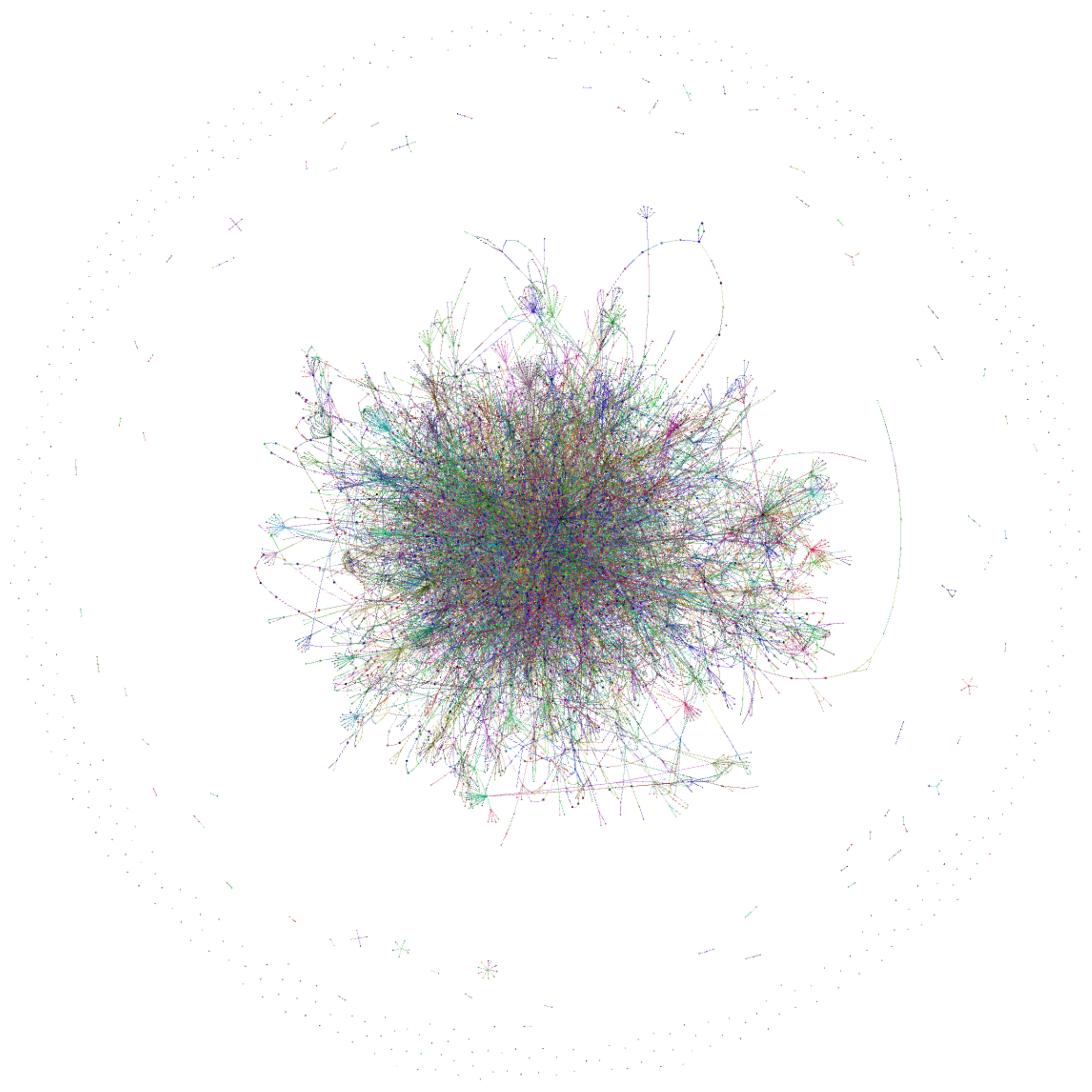} \\
        (c) MiniRAG & (d) TagRAG \\
    \end{tabular}
    \caption{Visualization of knowledge graph construction with Qwen3-4B on UltraDomain Mix.}
    \label{fig:visulaization}
\end{figure}

% To better distinguish the methods, we visualize their knowledge graphs on UltraDomain Mix in Figure~\ref{fig:visulaization}.
% Without a strong semantic parser, LightRAG with a smaller model extracts sparse entities and loose relations, yielding low-density graphs.
% GraphRAG and MiniRAG produce similar graphs. While GraphRAG adds global knowledge via community summaries, MiniRAG is more effective on small models through tailored retrieval.
% Leveraging hierarchical links, TagRAG builds well-connected graphs with high knowledge density and broad coverage.
To distinguish the methods, we visualize their knowledge graphs on UltraDomain Mix in Figure~\ref{fig:visulaization}.
Without a strong semantic parser, LightRAG with a smaller model extracts sparse entities and loose relations, resulting in low-density graphs.
GraphRAG and MiniRAG produce similar graphs. While GraphRAG adds global knowledge via community summaries, MiniRAG is more effective on small models through tailored retrieval.
Leveraging hierarchical links, TagRAG builds well-connected graphs with high knowledge density and broad coverage.

\section{Conclusion}
We propose TagRAG, a tag-guided hierarchical knowledge graph RAG framework that addresses key limitations of existing graph-based RAG systems, such as inefficient global reasoning and poor increment adaptability. By constructing domain-aware tag chains and enabling tag-guided graph retrieval, TagRAG achieves structured, fine-grained, and scalable knowledge integration. Our design reduces reliance on large language models and supports efficient construction and inference. The experimental results on the UltraDomain benchmark show that TagRAG achieves a 78.36\% average winning rate over baselines while offering significant gains in construction (14.6×) and retrieval (1.9×) efficiency compared to GraphRAG.

% \newpage

% \section{Limitations}
\section*{Limitations}
The indexing pipeline of TagRAG depends on LLM calls for object tag extraction and domain chain construction, which raises questions about cost, reproducibility, and robustness in fully automated scenarios.
In addition, TagRAG cannot handle multimedia data such as pictures and videos, which limits its application scope.

\section*{Acknowledgments}
The authors would like to thank the anonymous reviewers for their insightful comments. This work is supported by the National Key Research \& Develop Plan (Project No.2023YFF0725100) and Natural Science Foundation of China (Project No.U23A20298).

\bibliography{refs}

% \newpage
\clearpage

\appendix

\section{Detailed Experimental Settings}
\label{app:experimental details}

\subsection{Experimental Environment}
\label{app:environment}
In Table~\ref{tab:experimental environment}, we list the relevant experimental environment information, including hardware and software.

\begin{table}[h]
\centering
\small
\begin{tabular}{ll}
\toprule
& Configuration \\ \midrule
CPU & Intel(R) Xeon(R) Gold 6330 CPU \\ & @ 2.00GHz \\
GPU & NVIDIA RTX A6000 48GB \\
RAM & 256GB \\
Operating System & CentOS Linux 7 \\
CUDA & 12.4 \\
Python & 3.10.16 \\
\bottomrule
\end{tabular}
\caption{Experimental environment.}
\label{tab:experimental environment}
\end{table}

\subsection{Dataset Details}
\label{app:dataset}
We conduct experiments on four datasets, namely Agriculture, CS, Legal and Mix.
Agriculture, CS and Legal describe expertise in their respective fields, while Mix mixes knowledge from 18 fields.
As shown in Table~\ref{tab:dataset details}, the Agriculture and CS datasets are comparable in size, with Legal being the largest and Mix being the smallest.

\begin{table}[h]
\centering
\small
% \textwidth 表示自适应页面宽度，X 列会自动拉伸填充剩余空间
% 定义5列，第1列为普通左对齐，后4列为X类型（自适应宽度）
\begin{tabularx}{0.48\textwidth}{lXXXX}
\toprule
          & Agri  & CS     & Legal  & Mix    \\ \midrule
Docs      & 12    & 10     & 94     & 61     \\
Chunks    & 1756  & 1858   & 4294   & 579    \\
Tokens    & 2,017,886 & 2,306,535 & 5,081,069 & 619,009  \\
Size      & 8.56MB & 8.51MB  & 21.24MB & 2.54MB   \\ 
\bottomrule
\end{tabularx}
\caption{Dataset details.}
\label{tab:dataset details}
% 注：label建议使用无空格命名，如 tab:dataset_details，兼容性更好
% \label{tab:dataset_details}
\end{table}

% \begin{table}[h]
% \centering
% \small
% \begin{tabular}{lllll}
% \toprule
%  & Agri & CS & Legal & Mix \\ \midrule
% Docs & 12 & 10 & 94 & 61 \\
% Chunks & 1756 & 1858 & 4294 & 579 \\
% Tokens & 2,017,886 & 2,306,535 & 5,081,069 & 619,009 \\
% Size & 8.56MB & 8.51MB & 21.24MB & 2.54MB \\ 
% \bottomrule
% \end{tabular}
% \caption{Dataset Details.}
% \label{tab:dataset details}
% \end{table}

\subsection{Baselines}
\label{app:baseline}
To validate the performance and efficiency of global question answering, we compare TagRAG with the raw generation of three LLMs with different parameter amounts and the following four RAG-based methods:

\begin{itemize}
    \item \textbf{NaiveRAG}~\cite{lewis2020retrieval}: NaiveRAG, focusing on local information, combines LLMs with external knowledge retrieval to enhance response accuracy and relevance by dynamically fetching context from databases or documents before generating answers.
    \item \textbf{GraphRAG}~\cite{edge2024local}: GraphRAG builds an entity graph with community summaries, retrieving and aggregating partial summary responses into final answers. It grasps a complete global view and is capable of generating synthesized responses.
    % GraphRAG constructs a two-stage graph index: first building an entity knowledge graph, then pre-generating community summaries. Queries trigger partial responses from summaries, which are aggregated into a final answer. It grasps a complete global view and is capable of generating synthesized responses.
    % By extracting entities and relationships and delineating community summaries, GraphRAG builds documents into a structured knowledge graph and retrieves summaries to generate global answers.
    \item \textbf{LightRAG}~\cite{guo2024lightrag}: LightRAG combines graph structures with text indexing for dual-level retrieval, enabling fast, relevant knowledge access and real-time updates in dynamic environments. It is compared as a representative of lightweight.
    % LightRAG integrates graph structures with text indexing for dual-level retrieval, enhancing both low- and high-level knowledge discovery. It combines graph and vector representations for fast, relevant entity and relationship retrieval. An incremental update algorithm ensures real-time data integration for dynamic environments. It is compared as a representative of lightweight.
    % LightRAG integrates graph structures with text indexing for dual-level retrieval, improving low/high-level knowledge discovery. It combines graph and vector representations for efficient entity/relationship retrieval, boosting speed and relevance. An incremental update algorithm enables real-time data integration, ensuring adaptability in dynamic environments.
    \item \textbf{MiniRAG}~\cite{fan2025minirag}: MiniRAG is an efficient RAG system using semantic-aware graph indexing and lightweight topology-based retrieval for effective knowledge discovery with minimal semantic processing. As a comparison, it has excellent Graph-based RAG performance with small language models.
    % MiniRAG is a simple, efficient RAG system that features semantic-aware graph indexing and lightweight topology-enhanced retrieval, minimizing complex semantic processing while enabling effective knowledge discovery. As a comparison, it has excellent Graph-based RAG performance with small language models.
\end{itemize}

\subsection{Predefined Root Domain Tag Details}
In Table~\ref{tab:root domains}, we list the root domains with their descriptions for the Domain Tag Chain Organization on the UltraDomain datasets.

\begin{table}[h]
\centering
% \small
% \begin{tabular}{ll}
\begin{tabularx}{0.48\textwidth}{p{1.1cm}|X}
\toprule
Domain & Description \\ \midrule
Agri & Agriculture: Agriculture is the cultivation of crops and rearing of animals for food, fiber, and other products to sustain human life. \\
CS & Computer Science: Computer Science is the study of computational systems and algorithms, focusing on designing, analyzing, and applying software and hardware to solve complex problems. \\
Legal & Legal: Legal refers to anything pertaining to law, its principles, regulations, or the formal administration of justice within a society. \\
Mix & All disciplines: All disciplines refers to the complete range of academic, professional, and  practical fields of study, encompassing the humanities, sciences, arts, social sciences, technologies, and applied domains that collectively constitute human knowledge and activity. \\
\bottomrule
\end{tabularx}
\caption{Root domains with their descriptions.}
\label{tab:root domains}
\end{table}

\section{Extended experiments}

\subsection{Performance Analysis with glm-edge-1.5b-chat}

In order to verify the sensitivity of different methods to LLMs, we conducted further tests based on glm-edge-1.5b-chat~\footnote{\url{https://huggingface.co/zai-org/glm-edge-1.5b-chat}}.
As seen in Table~\ref{tab:main results glm-edge-1.5b-chat}, GraphRAG, LightRAG and MiniRAG, that perform quite well with qwen3-4b, exhibit catastrophic global reasoning.
This is due to the fact that both community summarization and path inference rely heavily on the capabilities of LLMs, causing these methods to perform poorly with lightweight models.
On the contrary, in this low resource situation, NaiveRAG still retains basic reasoning capabilities.
It can be seen that TagRAG can adapt to large language models with different parameter quantities and always has strong global reasoning capabilities.

\begin{table}[h]
\centering
\small
% \begin{tabular}{lrrrrr}
\begin{tabularx}{0.48\textwidth}{lXXXXX}
\toprule
& Agri& CS& Legal& Mix & Avg\\ \midrule
\multicolumn{6}{c}{\textit{v.s. NaiveRAG}}\\ \midrule
Comprehensiveness & 56.8& 73.3& 63.7& 53.3 & 61.8 
\\
Diversity & 57.5& 76.4& 74.8& 52.0 & 65.2 
\\
Empowerment & 64.8& 72.5& 64.7& 54.8 & 64.2 
\\
Overall & 60.8& 72.9& 60.0& 54.3 & 62.0 \\ \midrule
\multicolumn{6}{c}{\textit{v.s. GraphRAG}}\\ \midrule
Comprehensiveness & 100.0& 98.1& 96.9& 78.1 & 93.3 
\\
Diversity & 100.0& 98.3& 97.3& 79.2 & 93.7 
\\
Empowerment & 99.9& 98.5& 96.3& 78.1 & 93.2 
\\
Overall & 99.9& 98.3& 96.0& 78.3 & 93.1 \\ \midrule
\multicolumn{6}{c}{\textit{v.s. LightRAG}}\\ \midrule
Comprehensiveness & 100.0& 99.2& 96.7& 82.9 & 94.7 
\\
Diversity & 100.0&  99.6& 97.6& 84.1 & 95.3 
\\
Empowerment & 100.0& 99.2& 96.7& 80.5 & 94.1 
\\
Overall & 100.0& 99.2& 97.1& 82.1 & 94.6 \\ \midrule
\multicolumn{6}{c}{\textit{v.s. MiniRAG}}\\ \midrule
Comprehensiveness & 99.7& 99.2& 94.9& 79.7 & 93.4 
\\
Diversity & 98.3& 99.2& 96.0& 82.1 & 93.9 
\\
Empowerment & 99.6& 99.2& 94.7& 79.2 & 93.2 
\\
Overall & 99.7& 99.2& 94.7& 79.7 & 93.3 \\
\bottomrule
\end{tabularx}
% \end{tabular}
\caption{Main results: winning rates (\%) of TagRAG v.s. baselines with glm-edge-1.5b-chat across four datasets.}
\label{tab:main results glm-edge-1.5b-chat}
\end{table}

\begin{table}[t]
\centering
\begin{tabular}{lr}
\toprule
& \textit{v.s. GraphRAG} \\ \midrule
Comprehensiveness & 68.5\\
Diversity & 68.7\\
Empowerment & 66.5\\
Overall & 67.7\\  
\bottomrule
\end{tabular}
\caption{Lightweight adaption analysis: winning rates (\%) of TagRAG with Qwen3-1.7B v.s. GraphRAG with Qwen3-4B on CS.}
\label{tab:lightweight adaption analysis}
\end{table}

\subsection{Lightweight Adaption Analysis}
To illustrate the suitability and advantages of TagRAG on smaller LLMs, we use Qwen3-1.7B as a backbone for knowledge graph construction and inference, and compare it with other methods with Qwen3-4B in Table \ref{tab:lightweight adaption analysis}.
Even with a 57.5\% smaller LLM size, TagRAG absolutely dominates over NaiveRAG, GraphRAG and LightRAG and has comparable results to MiniRAG.
This suggests that the connection of domain chains and the fusion of domain-centric knowledge can be adapted to low-resource scenarios. TagRAG liberates the dependence on LLMs for graph construction.

\subsection{Semantic Inheritance Analysis}
To validate the effectiveness of the DAG-based domain tag chains in aggregating entity semantics and relationships, we conducted quantitative analyses on information retention and semantic inheritance.

First, to evaluate whether entity information propagates to higher levels, we calculated the cosine similarity between object tag descriptions and their upper-level domain tag summaries across different depths. As shown in Table~\ref{tab:info_propagation}, the average cosine similarity starts at 0.6823 for Level 1 and gradually converges to approximately 0.5 after Level 4. This indicates that domain tags closer to the objects retain more specific information, while higher-level tags effectively preserve a significant portion of the entity semantics, facilitating semantic clustering within the hierarchy.

\begin{table}[h]
\centering
\begin{tabular}{lc}
\toprule
Level & Cosine Sim Avg \\
\midrule
1 & 0.6823 \\
2 & 0.5806 \\
3 & 0.5305 \\
4 & 0.4968 \\
5 & 0.4919 \\
6 & 0.4927 \\
7 & 0.4980 \\
8 & 0.5092 \\
9 & 0.5166 \\
\bottomrule
\end{tabular}
\caption{Cosine similarity for information propagation}
\label{tab:info_propagation}
\end{table}

Second, to assess the stability of semantic integration between adjacent levels, we measured the similarity between high-level summaries and the concatenated descriptions of their immediate lower-level tags. The results, presented in Table~\ref{tab:adjacent_similarity}, show consistently high similarity scores averaging around 0.8 (peaking at 0.8509 for Level 4). This high degree of semantic consistency between neighboring levels confirms that the DAG-based structure ensures stable and effective inheritance of meaning throughout the tag chain.

\begin{table}[h]
\centering
\begin{tabular}{lc}
\toprule
Depth & Cosine Sim Avg \\
\midrule
1 (root) & 0.7916 \\
2 & 0.8248 \\
3 & 0.8478 \\
4 & 0.8509 \\
5 & 0.8381 \\
6 & 0.8341 \\
7 & 0.8378 \\
8 & 0.7427 \\
\bottomrule
\end{tabular}
\caption{Cosine similarity for adjacent level inheritance}
\label{tab:adjacent_similarity}
\end{table}

\subsection{Evaluation Metric Consistency Analysis}

To validate the coherence of our evaluation framework, we conducted a consistency analysis between the individual dimension metrics (Comprehensiveness, Diversity, and Empowerment) and the holistic Overall score on the UltraDomain-cs dataset. As presented in Table~\ref{tab:consistency_results}, all three metrics show high alignment with the final Overall judgment. Specifically, Empowerment demonstrates the highest consistency at 96.40\%, followed by Comprehensiveness (92.53\%) and Diversity (84.53\%). This strong correlation indicates that the evaluation model's scoring logic is robust and internally logical. Consequently, these results suggest that the Overall ranking can be effectively determined by aggregating the assessments from these three specific dimensions, reinforcing the validity of our multi-faceted evaluation approach.

\begin{table}[h]
\centering
\small
\begin{tabular}{lc}
\toprule
\textbf{Consistency Pair} & \textbf{Consistency Ratio} \\
\midrule
Comprehensiveness and Overall & 92.53\% \\
Diversity and Overall & 84.53\% \\
Empowerment and Overall & 96.40\% \\
\bottomrule
\end{tabular}
\caption{Consistency analysis between dimension metrics and overall score}
\label{tab:consistency_results}
\end{table}

\subsection{LLM-as-a-Judge Consistency Analysis}
\label{subsec:judge_consistency}

We further analyze the inter-consistency among these six evaluators on the UltraDomain-cs dataset. As shown in Table~\ref{tab:judge_consistency}, the results indicate a high degree of agreement. Specifically, approximately $64.1\%$ of the samples received identical scores from all six judges (6/6 consistency). When considering majority voting (agreement from at least 5 out of 6 judges), the consistency ratio rises to $82.4\%$, and it reaches $93.9\%$ when allowing for a wider margin (at least 4 out of 6). This high level of convergence validates the reliability of our evaluation outcomes.

\begin{table}[h]
\centering
\begin{tabular}{lc}
\toprule
\textbf{Consistency Level} & \textbf{Ratio (\%)} \\
\midrule
6/6 (Full Agreement) & 64.1 \\
$\geq$5/6 (Strong Majority) & 82.4 \\
$\geq$4/6 (Simple Majority) & 93.9 \\\bottomrule
\end{tabular}
\caption{Consistency analysis among six evaluators}
\label{tab:judge_consistency}
\end{table}

\subsection{Knowledge Graph Construction Robustness Analysis}
To validate the robustness of TagRAG, we construct a knowledge graph for the same document twice on the UltraDomain-cs dataset under the Qwen3-4B default settings (temperature=0.6, top-k=20, top-p=0.95). The results are presented in Table~\ref{tab:kg_construction}.
In the two knowledge graph constructions, there were a large number of overlapping object tags and tag chains. And, some object tags are different simply because they are represented separately, e.g., DATA- AND LEGAL-RELATED QUERIES are represented separately as DATA-RELATED QUERIES and LEGAL-RELATED QUERIES. In terms of entity description and summary, which TagRAG pays more attention to, the two graph constructions show a high degree of semantic similarity. This demonstrates the consistency and reliability of our approach in generating meaningful knowledge representations.

\begin{table*}[t]
\centering
\small
\begin{tabular}{lcl}
\toprule
Metric & Result & Explanation \\
\midrule
Object Tag Jaccard & 0.6735 & Jaccard similarity of object tags in two knowledge graphs \\
Object Tag Cosine & 0.9176 & Semantic similarity of descriptions for the same object tags \\
Tag Chain Jaccard & 0.7396 & Jaccard similarity of tag chains in two knowledge graphs \\
Summary Cosine & 0.9229 & Semantic similarity of domain tag summaries in two knowledge graphs \\
\bottomrule
\end{tabular}
\caption{Knowledge graph construction robustness analysis}
\label{tab:kg_construction}
\end{table*}

\subsection{Statistics of Knowledge Graph Construction}
We report the scale of knowledge graphs constructed by TagRAG and the baselines, as shown in Table~\ref{tab:statistics of knowledge graph construction}.
LightRAG extracts the least number of entities and relationships, resulting in its weakest knowledge representation.
GraphRAG and MiniRAG capture huge amounts of entities, but the relational connections are not as adequate as TagRAG. This makes the knowledge of the graphs they construct insufficiently connected to produce a global view.

\begin{table}[h]
\centering
\small
\begin{tabular}{l|llll}
    \toprule
    & Agri & CS & Legal & Mix \\ \midrule
    \multicolumn{5}{c}{NaiveRAG} \\ \midrule
    Docs & 12 & 10 & 94 & 61 \\
    Chunks & 1756 & 1858 & 4294 & 579 \\ \midrule
    \multicolumn{5}{c}{GraphRAG} \\ \midrule
    Entities & 38059 & 29803 & 35817 & 14393 \\
    Relationships & 33781 & 27598 & 49163 & 11114 \\ \midrule
    \multicolumn{5}{c}{LightRAG} \\ \midrule
    Entities & 19153 & 18737 & 19634 & 7592 \\
    Relationships & 911 & 2787 & 4008 & 266 \\ \midrule
    \multicolumn{5}{c}{MiniRAG} \\ \midrule
    Entities & 42777 & 35679 & 45381 & 16398 \\
    Relationships & 22218 & 20272 & 37816 & 9443 \\ \midrule
    \multicolumn{5}{c}{TagRAG} \\ \midrule
    $\mathcal{V}_o$ & 16086 & 14768 & 18479 & 7663 \\
    $\mathcal{E}_o$ & 29778 & 21822 & 43076 & 7941 \\
    $\mathcal{V}_d$ & 2893 & 1869 & 2489 & 2227 \\
    $\mathcal{E}_d$ & 4333 & 2886 & 4257 & 3568 \\
    $\mathcal{E}_{od}$ & 11462 & 7585 & 11819 & 3967 \\
    Entities & 18979 & 16637 & 20968 & 9890 \\
    Relationships & 45534 & 32258 & 59096 & 15421 \\ \midrule
    \bottomrule
\end{tabular}
\caption{Statistics of knowledge graph construction}
\label{tab:statistics of knowledge graph construction}
\end{table}

\subsection{Cross-domain Visual Analysis}

\begin{figure*}[t]
 \centering
 \includegraphics[width=0.98\linewidth]{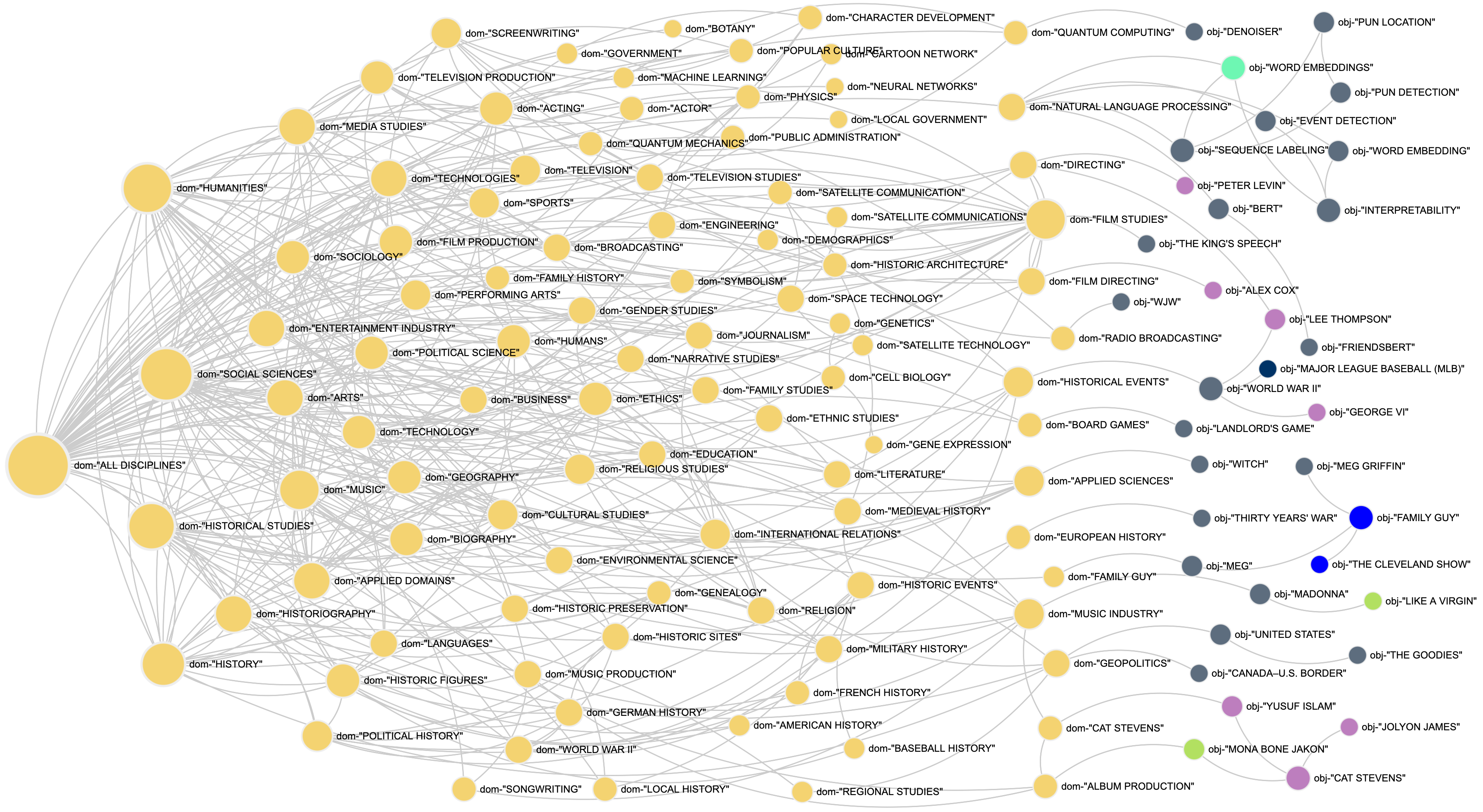}
 \centering
 \caption{Demonstration of tag chains on UltraDomain Mix.}
 \label{fig:cross-domain}
\end{figure*}

Leveraging the mapping of domain tag chains, TagRAG enables support for cross-domain scenarios. As illustrated in Figure~\ref{fig:cross-domain}, the predefined root domain tag ``\texttt{ALL DISCIPLINES}'' serves as the anchor for the entire domain knowledge graph, with diverse domains (e.g., social sciences, historical studies) interconnected in a hierarchical structure to guide cross-domain object tags. For instance, the object tag ``\texttt{THIRTY YEARS' WAR}'' links to the root domain via the chain ``\texttt{HISTORICAL STUDIES} - \texttt{EUROPEAN HISTORY}'', while another object tag ``\texttt{BERT}'' maps to ``\texttt{NATURAL LANGUAGE PROCESSING}'', which comes from a completely different field.
This demonstrates that TagRAG possesses the capability to integrate knowledge across different domains and adapt to cross-domain application scenarios.

\subsection{Case Study}

To intuitively reflect the effect of TagRAG, we demonstrate a query case for the UltraDomain CS corpus in Table~\ref{tab:TagTAG v.s. NaiveRAG} - \ref{tab:TagTAG v.s. MiniRAG}.
% ~\ref{tab:TagTAG v.s. NaiveRAG} to \ref{tab:TagTAG v.s. MiniRAG}.
In these cases, we can see that TagRAG is able to generate global analyses and responses, resulting in a complete lead in the four metrics of Comprehensiveness, Diversity, Empowerment, and Overall.

\section{Prompts}
\label{app:prompts}

We list the prompts involved in knowledge graph construction, retrieval enhancement generation, and experiments, including question generation, evaluation, object tag keyword extraction, domain tag chain organization, domain-centric knowledge fusion and tag knowledge-fused generation.

\begin{table*}[h]
\small
\begin{tabularx}{\textwidth}{p{2cm}|X}
    \toprule
    \multicolumn{1}{c|}{Query} & How can we select the most important features for the model using feature selection techniques? \\
    \hline
    
    \multicolumn{1}{c|}{TagRAG} & Selecting the most important features for a machine learning model is a critical step in the data preprocessing pipeline. It helps improve model performance, reduce overfitting, and enhance interpretability. Feature selection techniques can be broadly categorized into three types: \textbf{filter methods}, \textbf{wrapper methods}, and \textbf{embedded methods}. Each approach has its own advantages and is suitable for different scenarios depending on the problem, data, and model type. 
    
    \textbf{Filter Methods}

    Filter methods are the simplest and most computationally efficient techniques. They evaluate the relevance of features based on statistical measures and do not consider the model structure. These methods are ...
    
    \textbf{Wrapper Methods}
    
    Wrapper methods evaluate the importance of features by training and testing different subsets of features. These methods are more computationally intensive than filter methods but can lead to better ...
    
    \textbf{Embedded Methods}
    
    Embedded methods incorporate feature selection during the model training process. These methods are model-specific and are often used in regularization-based models. ...
    
    \textbf{Practical Considerations}
    
    1. Domain Knowledge: ... 2. Data Quality: ... 3. Computational Resources: ... 4. Model Type: ...
    
    In summary, feature selection is a crucial step in the machine learning pipeline that helps in building more efficient, accurate, and interpretable models. The choice of method depends on the specific problem, data characteristics, and computational resources available. By carefully selecting the most important features, we can significantly improve the performance of our models. \\
    
    \hline
    
    \multicolumn{1}{c|}{NaiveRAG} &  Feature selection is a crucial step in building effective machine learning models, as it helps to identify the most relevant variables that contribute to the model's predictive performance while reducing complexity and improving interpretability. The document outlines several feature selection techniques and their importance in the context of model training and evaluation.
    
    One of the key approaches mentioned is \textbf{backward selection}, which involves starting with all features and iteratively removing the least important ones until a stopping criterion is met. This method is particularly useful for reducing the number of features while maintaining model performance. ...
    
    In addition to backward selection, the document discusses the use of \textbf{multi-objective optimization} strategies, which aim to balance the trade-off between model performance and feature reduction. ...
    
    The document also provides an example of feature selection in the context of the \textbf{wine quality dataset}, where the goal is to predict wine quality based on ...
    
    In summary, feature selection techniques such as backward selection, multi-objective optimization, and the use of advanced optimization algorithms are essential for identifying the most important features for a model. These techniques not only improve model performance but also enhance the interpretability and efficiency of the model.\\ 
    
    \hline
    
    \multicolumn{1}{c|}{Evaluation} & \textbf{``Comprehensiveness''}: \{ ``Winner'': ``Answer 1 (TagRAG)'', ``Explanation'': ``Answer 1 provides a detailed and well-rounded overview of various feature selection techniques, covering filter methods, wrapper methods, and embedded methods. It includes specific examples such as Pearson’s correlation coefficient, mutual information, RFE, and Lasso regression, among others. In contrast, Answer 2 focuses primarily on backward selection and multi-objective optimization, providing less overall coverage of the different types of feature selection methods.'' \},
    
    \textbf{``Diversity''}: \{ ``Winner'': ``Answer 1 (TagRAG)'', ``Explanation'': ``Answer 1 presents a variety of techniques across different categories (filter, wrapper, and embedded), with examples from both statistical and machine learning methods. It also includes practical considerations like domain knowledge and computational resources. Answer 2 is more narrowly focused on backward selection and multi-objective optimization, offering fewer perspectives on feature selection.'' \},
    
    \textbf{``Empowerment''}: \{ ``Winner'': ``Answer 1 (TagRAG)'', ``Explanation'': ``Answer 1 gives readers a deeper understanding of the trade-offs and applications of different feature selection methods, such as the computational resources required and the types of models they suit best. It also includes practical tips, such as the importance of domain knowledge. This enables the reader to make more informed decisions about which technique to use. Answer 2, while informative, focuses more on specific methods without fully empowering the reader to make broad judgments about feature selection.'' \},
    
    \textbf{``Overall Winner''}: \{ ``Winner'': ``Answer 1 (TagRAG)'', ``Explanation'': ``Although Answer 2 is more direct, Answer 1 is the overall winner because it provides a more comprehensive, diverse, and empowering explanation. It covers a wider range of feature selection techniques and practical considerations, helping readers gain a thorough understanding of how to approach feature selection in different contexts. This broader perspective makes Answer 1 a more valuable resource for decision-making in feature selection.'' \} \\

    \bottomrule
\end{tabularx}
\caption{Case study: evaluation of TagTAG v.s. NaiveRAG with Qwen3-4B}
\label{tab:TagTAG v.s. NaiveRAG}
\end{table*}

\begin{table*}[h]
\small
\begin{tabularx}{\textwidth}{p{2cm}|X}
    \toprule
    \multicolumn{1}{c|}{Query} & How can we select the most important features for the model using feature selection techniques? \\
    \hline
    
    \multicolumn{1}{c|}{TagRAG} & Selecting the most important features for a machine learning model is a critical step in the data preprocessing pipeline. It helps improve model performance, reduce overfitting, and enhance interpretability. Feature selection techniques can be broadly categorized into three types: \textbf{filter methods}, \textbf{wrapper methods}, and \textbf{embedded methods}. Each approach has its own advantages and is suitable for different scenarios depending on the problem, data, and model type. 
    
    \textbf{Filter Methods}

    Filter methods are the simplest and most computationally efficient techniques. They evaluate the relevance of features based on statistical measures and do not consider the model structure. These methods are ...
    
    \textbf{Wrapper Methods}
    
    Wrapper methods evaluate the importance of features by training and testing different subsets of features. These methods are more computationally intensive than filter methods but can lead to better ...
    
    \textbf{Embedded Methods}
    
    Embedded methods incorporate feature selection during the model training process. These methods are model-specific and are often used in regularization-based models. ...
    
    \textbf{Practical Considerations}
    
    1. Domain Knowledge: ... 2. Data Quality: ... 3. Computational Resources: ... 4. Model Type: ...
    
    In summary, feature selection is a crucial step in the machine learning pipeline that helps in building more efficient, accurate, and interpretable models. The choice of method depends on the specific problem, data characteristics, and computational resources available. By carefully selecting the most important features, we can significantly improve the performance of our models. \\
    
    \hline
    
    \multicolumn{1}{c|}{GraphRAG} &  Feature selection techniques are essential for identifying the most important features for a model, as they help reduce dimensionality, improve model performance, and enhance interpretability. These techniques can be broadly categorized into three main types: filter methods, wrapper methods, and embedded methods. Filter methods evaluate features based on ... . Wrapper methods, on the other hand, use iterative model training to select ... . Embedded methods integrate feature selection directly into ... .
    
    The choice of technique depends on the specific problem, the type of data, and the goals of the modeling task. For instance, filter methods are computationally efficient and suitable for high-dimensional datasets, while wrapper methods may provide better performance but are more computationally intensive. Embedded methods are particularly useful when the model inherently performs feature selection, such as in regularization-based algorithms. Additionally, techniques like ... .
    
    The importance of feature selection lies in its ability to enhance model interpretability, reduce overfitting, and improve computational efficiency. By focusing on the most relevant features, models can achieve better accuracy and generalization, making them more effective for real-world applications. However, it is crucial to evaluate the impact of feature selection on model performance through validation techniques to ensure that the selected features contribute meaningfully to the model's predictive power. Ultimately, the most effective feature selection approach will depend on the specific characteristics of the dataset and the objectives of the modeling task. \\ 
    
    \hline
    
    \multicolumn{1}{c|}{Evaluation} & \textbf{``Comprehensiveness''}: \{ ``Winner'': ``Answer 1 (TagRAG)'', ``Explanation'': ``Answer 1 offers a highly detailed breakdown of the three main feature selection techniques—filter, wrapper, and embedded methods—with specific sub-methods under each category (e.g., Pearson correlation, RFE, Lasso). It also includes practical considerations such as domain knowledge, data quality, and model type, which extend the depth of coverage beyond just naming methods. Answer 2, while accurate, is more concise and general.''\},
    
    \textbf{``Diversity''}: \{ ``Winner'': ``Answer 1 (TagRAG)'', ``Explanation'': ``Answer 1 presents a richer variety of perspectives and techniques, covering a broader spectrum of tools within each method type and touching on ensemble models like Random Forest, as well as XGBoost and LightGBM. It also discusses non-technical factors (e.g., domain knowledge, computational resources), whereas Answer 2 is more method-focused and omits these complementary angles.''\},
    
    \textbf{``Empowerment''}: \{ ``Winner'': ``Answer 1 (TagRAG)'', ``Explanation'': ``Answer 1 is more empowering as it equips the reader with both theoretical knowledge and practical guidance. It explains when and why to use each method, provides context on computational cost, and suggests model-specific strategies. Answer 2 emphasizes general insights but lacks actionable detail for applying the techniques effectively.''\},
    
    \textbf{``Overall Winner''}: \{ ``Winner'': ``Answer 1 (TagRAG)', ``Explanation'': ``Answer 1 is the overall winner due to its superior comprehensiveness, greater diversity of content, and better empowerment of the reader. While Answer 2 is more direct, the depth and breadth of Answer 1 make it more valuable for users looking to understand and apply feature selection techniques effectively''\} \\

    \bottomrule
\end{tabularx}
\caption{Case study: evaluation of TagTAG v.s. GraphRAG with Qwen3-4B}
\label{tab:TagTAG v.s. GraphRAG}
\end{table*}

\begin{table*}[h]
\small
\begin{tabularx}{\textwidth}{p{2cm}|X}
    \toprule
    \multicolumn{1}{c|}{Query} & How can we select the most important features for the model using feature selection techniques? \\
    
    \hline
    
    \multicolumn{1}{c|}{TagRAG} & Selecting the most important features for a machine learning model is a critical step in the data preprocessing pipeline. It helps improve model performance, reduce overfitting, and enhance interpretability. Feature selection techniques can be broadly categorized into three types: \textbf{filter methods}, \textbf{wrapper methods}, and \textbf{embedded methods}. Each approach has its own advantages and is suitable for different scenarios depending on the problem, data, and model type. 
    
    \textbf{Filter Methods}

    Filter methods are the simplest and most computationally efficient techniques. They evaluate the relevance of features based on statistical measures and do not consider the model structure. These methods are ...
    
    \textbf{Wrapper Methods}
    
    Wrapper methods evaluate the importance of features by training and testing different subsets of features. These methods are more computationally intensive than filter methods but can lead to better ...
    
    \textbf{Embedded Methods}
    
    Embedded methods incorporate feature selection during the model training process. These methods are model-specific and are often used in regularization-based models. ...
    
    \textbf{Practical Considerations}
    
    1. Domain Knowledge: ... 2. Data Quality: ... 3. Computational Resources: ... 4. Model Type: ...
    
    In summary, feature selection is a crucial step in the machine learning pipeline that helps in building more efficient, accurate, and interpretable models. The choice of method depends on the specific problem, data characteristics, and computational resources available. By carefully selecting the most important features, we can significantly improve the performance of our models. \\
    
    \hline
    
    \multicolumn{1}{c|}{LightRAG} &  To select the most important features for a model using feature selection techniques, you can follow these steps based on the information provided in the Knowledge Base:
    
    1. \textbf{Understand the Purpose of Feature Selection}: Feature selection is the process of selecting ... .
    
    2. \textbf{Use Feature Selection Techniques}: There are several feature selection techniques that can  ... .
    
    3. \textbf{Apply Feature Selection Algorithms}: The Knowledge Base mentions that feature selection can ... .
    
    4. \textbf{Evaluate Feature Importance}: After selecting the features, it is important to evaluate ... .
    
    5. \textbf{Integrate Feature Selection with Model Training}: Once the relevant features are selected, they ... .
    
    6. \textbf{Validate the Selected Features}: Finally, the selected features should be validated using ... .
    
    By following these steps, you can effectively select the most important features for your model using feature selection techniques, leading to improved model performance and interpretability. \\ 
    
    \hline
    
    \multicolumn{1}{c|}{Evaluation} & \textbf{``Comprehensiveness''}: \{``Winner'': ``Answer 1 (TagRAG)'', ``Explanation'': ``Answer 1 provides a highly detailed breakdown of the different feature selection methods (filter, wrapper, and embedded), including specific techniques and algorithms within each category. It offers a thorough explanation of statistical measures, domain knowledge considerations, and computational resources, making it more comprehensive in covering all the necessary aspects of feature selection. In contrast, Answer 2 is relatively high-level and misses some of the specific techniques and broader considerations like the different types of feature selection methods.''\},
    
    \textbf{``Diversity''}: \{``Winner'': ``Answer 1 (TagRAG)'', ``Explanation'': ``Answer 1 offers a wide range of techniques and insights into feature selection methods, discussing multiple algorithms within each category (e.g., Pearson's correlation, Recursive Feature Elimination, Lasso, Random Forest) and highlighting their respective strengths and applications. It also touches on practical aspects like domain knowledge and computational resources. On the other hand, Answer 2 presents fewer perspectives and sticks more closely to general steps, without delving into diverse methodologies or offering a variety of relevant examples.''\},
    
    \textbf{``Empowerment''}: \{``Winner'': ``Answer 1 (TagRAG)'', ``Explanation'': ``Answer 1 empowers the reader by explaining the different techniques in detail and providing concrete examples (e.g., Pearson’s correlation, RFE, Lasso, and Random Forest). It helps the reader understand how to apply each method and why one might choose one over another based on data, model type, and resources. It provides enough depth to make informed judgments. Answer 2, while giving a process-oriented approach, doesn’t provide enough detail on how to evaluate or apply these methods in practice, which reduces the reader's ability to fully grasp how to make decisions about feature selection.''\},
    
    \textbf{``Overall Winner''}: \{``Winner'': ``Answer 1 (TagRAG)'', ``Explanation'': ``Answer 1 is the overall winner due to its comprehensive coverage, diversity of insights, and empowerment of the reader with clear examples and practical considerations. It offers a far more detailed and multifaceted view of feature selection techniques, enabling the reader to make informed decisions about which approach to use in various scenarios. While Answer 2 is more direct and streamlined, it lacks the depth and richness necessary to provide a full understanding of the topic.''\} \\

    \bottomrule
\end{tabularx}
\caption{Case study: evaluation of TagTAG v.s. LightRAG with Qwen3-4B}
\label{tab:TagTAG v.s. LightRAG}
\end{table*}

\begin{table*}[h]
\small
\begin{tabularx}{\textwidth}{p{2cm}|X}
    \toprule
    \multicolumn{1}{c|}{Query} & How can we select the most important features for the model using feature selection techniques? \\
    
    \hline
    
    \multicolumn{1}{c|}{TagRAG} & Selecting the most important features for a machine learning model is a critical step in the data preprocessing pipeline. It helps improve model performance, reduce overfitting, and enhance interpretability. Feature selection techniques can be broadly categorized into three types: \textbf{filter methods}, \textbf{wrapper methods}, and \textbf{embedded methods}. Each approach has its own advantages and is suitable for different scenarios depending on the problem, data, and model type. 
    
    \textbf{Filter Methods}

    Filter methods are the simplest and most computationally efficient techniques. They evaluate the relevance of features based on statistical measures and do not consider the model structure. These methods are ...
    
    \textbf{Wrapper Methods}
    
    Wrapper methods evaluate the importance of features by training and testing different subsets of features. These methods are more computationally intensive than filter methods but can lead to better ...
    
    \textbf{Embedded Methods}
    
    Embedded methods incorporate feature selection during the model training process. These methods are model-specific and are often used in regularization-based models. ...
    
    \textbf{Practical Considerations}
    
    1. Domain Knowledge: ... 2. Data Quality: ... 3. Computational Resources: ... 4. Model Type: ...
    
    In summary, feature selection is a crucial step in the machine learning pipeline that helps in building more efficient, accurate, and interpretable models. The choice of method depends on the specific problem, data characteristics, and computational resources available. By carefully selecting the most important features, we can significantly improve the performance of our models. \\
    
    \hline
    
    \multicolumn{1}{c|}{MiniRAG} &  Selecting the most important features for a model is a crucial step in the machine learning pipeline, as it can significantly impact the model's performance, interpretability, and efficiency. Feature selection techniques help ... . Here's how you can approach feature selection using various techniques:

    1. \textbf{Understanding the Problem and Data}
    Before applying any feature selection techniques, it's  ... .
    
    2. \textbf{Feature Selection Techniques}
    There are several feature selection techniques, which can ... .
    
    3. \textbf{Feature Selection in Practice}
    To apply feature selection in practice, you can follow these steps: ... .
    
    4. \textbf{Example: Using Lasso Regression for Feature Selection}
    Here's an example of using Lasso Regression for feature selection in Python using ... .
    
    5. \textbf{Conclusion}
    Feature selection is a critical step in the machine learning pipeline. By selecting the most important features, you can improve the model's performance, reduce overfitting, and make the model more interpretable. The choice of feature selection technique depends on the specific problem, the size of the dataset, and the computational resources available. \\ 
    
    \hline
    
    \multicolumn{1}{c|}{Evaluation} & \textbf{``Comprehensiveness''}: \{``Winner'': ``Answer 1 (TagRAG)'', ``Explanation'': ``Answer 1 is more comprehensive as it goes into greater detail about each feature selection technique, explaining the methods in-depth, including the differences between various types of techniques (filter, wrapper, and embedded methods). It also discusses practical considerations and highlights the importance of domain knowledge, data quality, and model type, which adds value and depth to the response.''\},
    
    \textbf{``Diversity''}: \{``Winner'': ``Answer 1 (TagRAG)'', ``Explanation'': ``Answer 1 presents a wider variety of feature selection techniques, offering diverse methods within each category (filter, wrapper, and embedded). It also discusses additional considerations like computational resources, model type, and domain knowledge, giving a broader range of insights on the topic.''\},
    
    \textbf{``Empowerment''}: \{``Winner'': ``Answer 1 (TagRAG)'', ``Explanation'': ``Answer 1 empowers the reader by thoroughly explaining each technique's use, advantages, and practical scenarios. It also provides valuable insights on how the selection method should align with the problem at hand, giving readers the tools to make informed decisions based on their specific context.''\},
    
    \textbf{``Overall Winner''}: \{``Winner'': ``Answer 1 (TagRAG)'', ``Explanation'': ``Although Answer 2 is more direct, Answer 1 provides a more comprehensive, diverse, and empowering response that thoroughly covers the topic, making it the overall winner. It offers in-depth explanations, a variety of methods, and detailed guidance, which would better equip the reader to understand and apply feature selection techniques in different scenarios.''\} \\

    \bottomrule
\end{tabularx}
\caption{Case study: evaluation of TagTAG v.s. MiniRAG with Qwen3-4B}
\label{tab:TagTAG v.s. MiniRAG}
\end{table*}

%%%%%%%%%%%%%%%%%%%%%%%%%%%%%%%%%%%%%%%%%%%%%%%%%%%%%%%%%%%%%%%%%%%%%%%%%%%%%%%%%%%%%%%%%%

% \newpage
\clearpage

\begin{figure*}[h]
\centering
\small
\begin{tcolorbox}[title=Prompt of question generation, colframe=gray, colback=white]
    Given the following description of a dataset: \{total\_description\}.
    Please identify 5 potential users who would engage with this dataset. For each user, list 5 tasks they would perform with this dataset. Then, for each (user, task) combination, generate 5 questions that require a high-level understanding of the entire dataset.
    Output the results in the following structure:\\
    - User 1: [user description]\\
        \hspace*{2em}- Task 1: [task description]\\
            \hspace*{4em}- Question 1:\\
            \hspace*{4em}...\\
            % \hspace*{4em}- Question 2:\\
            % \hspace*{4em}- Question 3:\\
            % \hspace*{4em}- Question 4:\\
            \hspace*{4em}- Question 5:\\
        % \hspace*{2em}- Task 2: [task description]\\
        \hspace*{2em}...\\
        \hspace*{2em}- Task 5: [task description]\\
    % - User 2: [user description]\\
        % \hspace*{2em}...\\
    ...\\
    - User 5: [user description]\\
        \hspace*{2em}...
\end{tcolorbox}
\end{figure*}

\begin{figure*}[h]
\centering
\small
\begin{tcolorbox}[title=Prompt of evaluation, colframe=gray, colback=white]
    ---Role---\\
    You are an expert tasked with evaluating two answers to the same question based on three criteria: **Comprehensiveness**, **Diversity**, **Empowerment**, and **Directness**.\\
    You will evaluate two answers to the same question based on three criteria: **Comprehensiveness**, **Diversity**, and **Empowerment**.\\
    - **Comprehensiveness**: How much detail does the answer provide to cover all aspects and details of the question?\\
    - **Diversity**: How varied and rich is the answer in providing different perspectives and insights on the question?\\
    - **Empowerment**: How well does the answer help the reader understand and make informed judgments about the topic?\\
    For each criterion, choose the better answer (either Answer 1 or Answer 2) and explain why. Then, select an overall winner based on these three categories.\\
    Here is the question: \{query\}\\
    % \{query\}\\
    Here are the two answers: \\
    **Answer 1:** \{answer1\}\\
    % \{answer1\}\\
    **Answer 2:** \{answer2\}\\
    % \{answer2\}\\

    Evaluate both answers using the three criteria listed above and provide detailed explanations for each criterion.
    Output your evaluation in the following JSON format:

    \{``Comprehensiveness'': \{``Winner'': ``[Answer 1 or Answer 2]'', ``Explanation'': ``[Provide explanation here]''\}, \\
        \hspace*{0.5em}``Diversity'': \{``Winner'': ``[Answer 1 or Answer 2]'', ``Explanation'': ``[Provide explanation here]''\}, \\
        \hspace*{0.5em}``Empowerment'': \{``Winner'': ``[Answer 1 or Answer 2]'', ``Explanation'': ``[Provide explanation here]''\}, \\
        \hspace*{0.5em}``Overall Winner'': \{``Winner'': ``[Answer 1 or Answer 2]'', ``Explanation'': ``[Summarize why this answer is the overall winner based on the three criteria]''\}\}

    % \{ \\
    %     \hspace*{2em}``Comprehensiveness'': \{ \\
    %         \hspace*{4em}``Winner'': ``[Answer 1 or Answer 2]'', \\
    %         \hspace*{4em}``Explanation'': ``[Provide explanation here]'' \\
    %     \hspace*{2em}\}, \\
    %     \hspace*{2em}``Diversity'': \{ \\
    %         \hspace*{4em}``Winner'': ``[Answer 1 or Answer 2]'', \\
    %         \hspace*{4em}``Explanation'': ``[Provide explanation here]'' \\
    %     \hspace*{2em}\}, \\
    %     \hspace*{2em}``Empowerment'': \{ \\
    %         \hspace*{4em}``Winner'': ``[Answer 1 or Answer 2]'', \\
    %         \hspace*{4em}``Explanation'': ``[Provide explanation here]'' \\
    %     \hspace*{2em}\}, \\
    %     \hspace*{2em}``Overall Winner'': \{ \\
    %         \hspace*{4em}``Winner'': ``[Answer 1 or Answer 2]'', \\
    %         \hspace*{4em}``Explanation'': ``[Summarize why this answer is the overall winner based on the three criteria]'' \\
    %     \hspace*{2em}\} \\
    % \}
\end{tcolorbox}
\end{figure*}

\begin{figure*}[h]
\centering
\small
\begin{tcolorbox}[title=Prompt of object tag keyword extraction, colframe=gray, colback=white]
    ---Goal--- \\
    Given a domain-specific text document and a list of keyword types, summarize keywords from the text and generate their relationships.
    Use \{language\} as output language. \\
    ---Steps--- \\
    1. Summarize keywords from the text. For each summarized keyword, generate the following information: \\
    - keyword\_name: Name of the keyword, use same language as input text. If English, capitalized the name. \\
    - keyword\_type: Type of the keyword that can classify the keyword. \\
    - keyword\_description: Comprehensive description of the keyword's attributes and activities \\
    Format each keyword as (``keyword''\{tuple\_delimiter\}$<keyword\_name>$\{tuple\_delimiter\}$<keyword\_type>$ \\
    \{tuple\_delimiter\}$<keyword\_description>$) \\
    2. From the keywords summarized in step 1, generate all pairs of (source\_keyword, target\_keyword) that are *clearly related* to each other.  \\
    Don't create source\_keyword or target\_keyword that are not summarized in step 1. \\
    For each pair of related keywords, generate the following information: \\
    - source\_keyword: name of the source keyword, as summarized in step 1 \\
    - target\_keyword: name of the target keyword, as summarized in step 1 \\
    - relationship\_description: explanation as to why you think the source keyword and the target keyword are related to each other \\
    Format each relationship as (``relationship''\{tuple\_delimiter\}$<source\_keyword>$\{tuple\_delimiter\} \\
    $<target\_keyword>$\{tuple\_delimiter\}$<relationship\_description>$) \\
    3. Return output in \{language\} as a single list of all the keywords and relationships generated in steps 1 and 2. Use **\{record\_delimiter\}** as the list delimiter. \\
    4. When finished, output \{completion\_delimiter\} \\
    
    ---Examples--- \\
    \{examples\} \\
    ---Real Data--- \\
    Text: \{input\_text\} \\
    Output:
\end{tcolorbox}
\end{figure*}

\begin{figure*}[h]
\centering
\small
\begin{tcolorbox}[title=Prompt of domain tag chain organization, colframe=gray, colback=white]
    ---Goal--- \\
    Given a domain tag with its description and an object tag with its description, generate the relationship chain between them. \\
    Use \{language\} as output language. \\
    ---Steps--- \\
    1. Generate the relationship chain between the domain tag and the object tag. Present all domain tags consisting of the following information: \\
    - domain\_name: Name of the domain, use same language as input text. If English, capitalized the name. \\
    - domain\_description: Comprehensive description of the domain tag. \\
    Format each domain tag as $<domain\_name>$\{explanation\_delimiter\}$<domain\_description>$ and connect domain tags with **\{inference\_delimiter\}**. \\
    2. Generate the relationship description between the object tag and the generated relationship chain in step 1. Use **\{tuple\_delimiter\}** as the delimiter. \\
    3. Return output in \{language\} as a single relationship chain generated in step 1 and a relationship description generated in step 2. \\
    4. When finished, output \{completion\_delimiter\} \\
    ---Examples--- \\
    \{examples\} \\
    ---Real Data--- \\
    Domain tag name: \{domain\_tag\_name\} \\
    Domain tag description: \{domain\_tag\_description\} \\
    Object tag name: \{object\_tag\_name\} \\
    Object tag description: \{object\_tag\_description\} \\
    Output:
\end{tcolorbox}
\end{figure*}

\begin{figure*}[h]
\centering
\small
\begin{tcolorbox}[title=Prompt of domain-centric knowledge fusion, colframe=gray, colback=white]
    ---Goal---
    
    Given a chain of domain tags with their descriptions, the summary of the chain, relevant object tags with their descriptions and the relationship descriptions, summarize the domain by injecting these object tags at a high level.
    Use \{language\} as output language. \\
    ---Domain tag chain--- \\
    \{domain\_tag\_chain\} \\
    ---Chain summary--- \\
    \{chain\_summary\} \\
    ---Object tags--- \\
    \{object\_tags\} \\
    ---relationships--- \\
    \{relationships\} \\
    Output:
\end{tcolorbox}
\end{figure*}

\begin{figure*}[h]
\centering
\small
\begin{tcolorbox}[title=Prompt of tag knowledge-fused generation, colframe=gray, colback=white]
    ---Role--- \\
    You are a helpful assistant responding to questions about data in the tables provided. \\
    
    ---Goal--- \\
    Generate a response of the target length and format that responds to the user's question, summarizing all information in the input data tables appropriate for the response length and format, and incorporating any relevant general knowledge. \\
    Do not include information where the supporting evidence for it is not provided. \\
    
    ---Target response length and format--- \\
    \{response\_type\} \\
    ---Data tables--- \\
    \{context\_data\} \\
    
    Add sections and commentary to the response as appropriate for the length and format. Style the response in markdown.
\end{tcolorbox}
\end{figure*}

\end{document}